\documentclass[aps,prd,reprint,10pt,showkeys,nofootinbib]{revtex4-2}

\usepackage{amssymb,amsmath,amsfonts,amsthm,wasysym}
\usepackage{amsbsy} 
\usepackage{epsfig}
\usepackage{latexsym}
\usepackage{comment}
\usepackage{tensor}
\usepackage{adjustbox}
\usepackage{graphicx}
\usepackage[]{cancel}
\usepackage{xcolor}
\usepackage{enumitem}
\usepackage{subcaption}
\usepackage{hyperref}
\usepackage{booktabs}

\makeatletter
\def\l@subsubsection#1#2{}
\makeatother

\makeatletter 

\renewcommand\onecolumngrid{% <<<<<<
	\do@columngrid{one}{\@ne}%
	\def\set@footnotewidth{\onecolumngrid}% <<<<<<<<<<<<<<<<
	\def\footnoterule{\kern-6pt\hrule width 1.5in\kern6pt}%
}

\renewcommand\twocolumngrid{% <<<<<<
	\def\footnoterule{% restore rule
		\dimen@\skip\footins\divide\dimen@\thr@@
		\kern-\dimen@\hrule width.5in\kern\dimen@}
	\do@columngrid{mlt}{\tw@}
}%

\makeatother    
\def\be{\begin{equation}}
\def\ee{\end{equation}}
\def\dd{{\rm d}}
\def\bes{\begin{eqnarray}}
\def\ees{\end{eqnarray}}

\DeclareMathOperator{\sgn}{sgn}

\DeclareMathOperator{\tr}{tr}
\begin{document}
	
\title{Generalised Gaussian states in group field theory and $\mathfrak{su(1,1)}$ quantum cosmology}
\author{Andrea Calcinari}
\email{acalcinari1@sheffield.ac.uk}
\author{Steffen Gielen}
\email{s.c.gielen@sheffield.ac.uk}
\affiliation{School of Mathematics and Statistics, University of Sheffield, Hicks Building, Hounsfield Road, Sheffield S3 7RH, United Kingdom}
\date{\today}

\begin{abstract}
We define generalised Gaussian states for quantum cosmological models based on the $\mathfrak{su(1,1)}$ algebra, with particular emphasis on its realisation in group field theory for a single field mode, and study their semiclassical properties. These states are generalisations of coherent, squeezed and thermal states considered previously. As two possible characterisations of semiclassicality, we contrast the requirement of small relative fluctuations in volume and energy with the saturation of the Robertson--Schr\"odinger uncertainty principle. We find that for the most general class of states the appearance of small relative fluctuations, which we take as the main criterion relevant for the emergence of cosmology, is mostly determined by the amount of displacement used to define the state. We also observe that defining such generalised Gaussian states is less straightforward in the algebraic approach to canonical quantisation of group field theory, and discuss special cases.
\end{abstract}

\maketitle

%----------------------------------------------------------------------------
\section{Introduction}
%----------------------------------------------------------------------------

Describing macroscopic phenomena in quantum mechanics and quantum field theory often requires a semiclassical approximation. In a canonical setting, such an approximation may be implemented by choosing suitable states with (to be specified) semiclassical properties, whereas in the path integral it is often associated with stationary phase approximations. A prime example, closely related to what we will discuss in the following, is the description of a macroscopic electromagnetic field in terms of coherent or squeezed states in quantum optics. In general, an important question is whether coherence or semiclassical properties of an initially chosen state will be preserved under time evolution. This is famously the case for the harmonic oscillator (or free quantum fields) but not for more general interacting quantum systems.

In quantum gravity and quantum cosmology, identifying a semiclassical spacetime description is a rather crucial requirement, both conceptually and for making the link to the low-energy world in which we do not observe spacetime superpositions. In traditional Wheeler--DeWitt quantum cosmology, semiclassical spacetime was often identified in a WKB (Wentzel--Kramers--Brillouin) regime in which the wavefunction is assumed to be highly oscillating. This approximation is at the heart of applications to cosmological perturbation theory, in which the curved spacetime quantum field theory setting of inflationary cosmology emerges from the semiclassical limit of quantum cosmology (see, e.g., \cite{HalliwellHawking,Kiefer1987,WKBReview}). The notion of semiclassicality applied here is different from that of using coherent states or wavepackets; a WKB state is (by assumption) not localised in configuration space, but rather describes an entire classical trajectory ``all at once''.

Loop quantum gravity (LQG), one of the most established approaches to the problem of quantum gravity, offers its own proposals for the semiclassical limit. Again, here we focus on the  canonical formulation of the theory, in which one works with quantum states living on superpositions of graphs. A class of coherent states for LQG, which has found many applications in the literature, was proposed in \cite{ThiemannCoh1,*ThiemannCoh2}, whereas more recent proposals include \cite{UNCoherent,*TwistedCoherent}. These states realise the traditional properties of coherent states, peakedness around a given classical configuration with small uncertainties. In general, given the rather complicated dynamics of full LQG, it is not clear whether these semiclassical properties would be preserved dynamically.

In this paper we focus on the group field theory (GFT) approach \cite{FreidelGFT,*Oriti_GFTandLQG} whose canonical formulation is closely related to the canonical formalism for LQG; Fock space quantisations of GFT lead to state spaces that can be interpreted in terms of spin-network states of LQG \cite{Oriti_GFT2ndLQG}. One may ask what kind of GFT quantum states could be used for a semiclassical, macroscopic limit of the theory, relevant in particular in the application to cosmology. Here a particularly influential idea has been to use analogies with condensed matter physics and think of a ``condensate'' of quanta of geometry, or LQG spin-network vertices \cite{GFTquantumST_Oriti,*ORITI2017235,Gielen_2016,GFTcosmoLONGpaper}. Such a condensate can be characterised in a mean-field approximation, or equivalently using a Fock coherent state built on the fundamental field operators or the annihilation and creation operators of the theory. Many cosmological applications of GFT have focused exclusively on such coherent states in extracting a semiclassical limit \cite{Oriti_2016,*BOriti_2017}. Our goal here is to broaden this perspective, and discuss a class of semiclassical states that go beyond the simplest choice of Fock coherent states. In particular, we want to write down the most general {\em Gaussian} state associated with a single GFT (Peter--Weyl) field mode, and also discuss mixed states following the work of \cite{Isha_thermalGFT,*Isha_thermalGFT2,*IshaDaniele} on thermal states in GFT. We will build on the work of \cite{Gielen_2020}, which already discussed some more general types of coherent states built on the $\mathfrak{su(1,1)}$ algebra of observables most relevant for cosmology, and extend the results of that paper substantially.

{While} we are mostly interested in GFT models, our results are much more generally applicable to any $\mathfrak{su(1,1)}$ cosmological scenario. For example, as pointed out in \cite{Bojo_2019}, isotropic models of loop quantum cosmology and (bosonic) GFT cosmology can be seen as different realisations of the same underlying structure (sometimes called ``harmonic cosmology'' \cite{HarC1, *HarC2}). The $\mathfrak{su(1,1)}$ Lie algebra was then also investigated in detail in the context of loop cosmology in \cite{su11LQC}, where it was realised that dynamics could be implemented as $SU(1,1)$ transformations, and it was later associated with the ``Complexifier-Volume-Hamiltonian (CVH) algebra'' in \cite{CVHalgebra} (see also \cite{Boden1, *Boden2, *Boden3} for more recent work).

When discussing the relative merits of possible choices of semiclassical states, we need to be clear about what properties we require for a state to be considered semiclassical. Here we follow to a large extent the criteria set out in the context of GFT in \cite{Gielen_2020}; our main requirement for semiclassicality is that the relative uncertainty in the volume, $(\Delta \hat{V})/\langle\hat{V}\rangle$, can be made arbitrarily small, in particular at late times or large volumes after dynamical evolution. This criterion is similar to what is often required for a semiclassical limit in loop quantum cosmology \cite{Taveras,*AshtekarGupt}. We also require a small relative uncertainty in the Hamiltonian (associated with the matter coupled to gravity in GFT), which is time-independent. In contrast, one could also require that a semiclassical state saturate the lower bound on uncertainties implied by the uncertainty principle, in its stronger Robertson--Schr\"odinger form. We will argue that this second requirement seems less relevant physically, since the right-hand side of the uncertainty principle is in general state-dependent and one can end up with an equality for which both sides are large. For the context of GFT and the most relevant cosmological observables, energy and volume, we will find a conflict between the two requirements: states with small relative uncertainties do not saturate the Robertson--Schr\"odinger inequality, while those that saturate the inequality do not have small relative uncertainties. This discrepancy was observed for squeezed states in \cite{Gielen_2020}; again we generalise this discussion to general Gaussian states.

We will show that while general Gaussian states can be constructed using displacement, squeezing and thermality, semiclassical properties are mostly determined only by the magnitude of displacement: squeezed or thermal states alone are not semiclassical in the sense we require, and hence do not lead to a good interpretation in terms of emergent cosmology. These results can be seen as justifying to an extent the emphasis on Fock coherent states in the GFT literature. While most of our analysis uses the deparametrised approach to the canonical quantisation, in which a scalar matter field is used as a time variable throughout \cite{relham_Wilson_Ewing_2019,*relhamadd}, we also discuss general Gaussian states in the more commonly used ``algebraic'' approach based on a kinematical Hilbert space. In that setting, we find that generalisations of simple coherent states are difficult to construct, and only very simple versions of squeezing and thermality can be straightforwardly defined. We also encounter a number of technical issues related to divergences in the definition of states and observables. Ignoring these as much as possible, the general qualitative statements agree with those found in the deparametrised approach.

Section \ref{gftcosmosection} reviews the main ingredients in the canonical quantisation of GFT, leading to the emergence of homogeneous and isotropic cosmology (satisfying a generalised Friedmann equation) from the simplest dynamics for a single field mode in the Peter--Weyl decomposition. In section \ref{StatesSection} we discuss different definitions of semiclassicality and study the examples of coherent and squeezed states explicitly. This analysis is then generalised to general Gaussian states in section \ref{GaussSection}. Given that the algebraic approach to canonical quantisation is used in most of the GFT literature, in section \ref{algebraicsection} we discuss our efforts at obtaining similar  types of states in that approach. Appendices contain details on expectation values, variances and covariances and the Robertson--Schr\"odinger uncertainty principle; the general definition of Gaussian states and the thermofield formalism; and details on the possible construction of semiclassical ``condensate''  states in the algebraic approach.

%----------------------------------------------------------------------------
\section{Cosmology from Group Field Theory}\label{gftcosmosection}
%----------------------------------------------------------------------------

GFT is a relatively young, nonperturbative and background-independent approach to quantum gravity, which generalises matrix and tensor models \cite{matrix,*ColourTensor} by including Lie group structures imported from formalisms like LQG and spin foam models \cite{PerezSFQG}. The fundamental object in this framework is the group field $\varphi$, whose arguments replace the discrete indices of a tensor with a number of continuous variables taking values in a Lie group. In models of interest for us, this group consists of four copies of $SU(2)$, so as to resemble the structure of spin networks of LQG, but different choices are possible (see, e.g., \cite{BCGFT}). One can then couple gravity to a free massless scalar field $\chi\in \mathbb{R}$ which can serve as a relational time variable \cite{Oriti_2016,*BOriti_2017}. A real group field is then a map
\begin{equation}\label{groupfield}
	\begin{aligned}
&	\varphi: SU(2)^4 \times \mathbb{R} \rightarrow \mathbb{R}\,,\\&  \varphi (g_I,\chi) = \varphi (g_I h,\chi) \quad \forall h\in SU(2)\,,
	\end{aligned}
\end{equation}
where requiring invariance of the field under the right diagonal group action provides a notion of discrete gauge invariance. The general action for a real group field reads 
\begin{equation}\label{GFTaction}
	S[\varphi] = \frac{1}{2}\int {\rm d}^4g\; {\rm d}^4g'\; {\rm d}\chi\;\varphi(g_I,\chi) K(g_I,g'_I)\varphi(g'_I,\chi) + V[\varphi]\,,
\end{equation}
where ${\rm d}g$ is the Haar measure on $SU(2)$. Here, $K(g_I,g'_I)$ is a quadratic kinetic operator and $V[\varphi]$ is a generally nonlocal interaction term. Requiring the kinetic term to respect the symmetries of a minimally coupled massless scalar field (shift and sign reversal symmetries) implies that $K(g_I,g'_I)$ should not depend on $\chi$ but be a differential operator in $\chi$, without derivatives of odd powers \cite{Oriti_2016,*BOriti_2017,Li_2017}. The simplest choice is therefore to assume the minimal form
\begin{equation}\label{kinetic}
	K (g_I,g'_I) = 	K^{(0)} (g_I,g'_I) + K^{(2)} (g_I,g'_I) \partial_\chi^2 \,,
\end{equation}
which is commonly adopted in the literature \cite{relham_Wilson_Ewing_2019,*relhamadd,Aniso,Marchetti:2022igl,*Marchetti:2022nrf}. Radiative corrections coming from renormalisation \cite{Geloun_2013} can dictate the specific forms of $K^{(0)}$ and $K^{(2)}$, but we can leave them general for our purposes. Within a broader class of models one can in principle have higher derivatives with respect to $\chi$ \cite{Oriti_2016,*BOriti_2017,Li_2017}; \eqref{kinetic} would then be seen as an approximation in which the contribution of these higher-derivative terms is small.

Similarly to what can be done in tensor models for quantum gravity, and thanks to the connection with spin foam models, a perturbative expansion of the GFT partition function can formally generate an infinite sum over discrete geometries, or Feynman graphs $h$. For a real $\varphi$, and for models with only a single interaction with coupling $\lambda$, one finds
\begin{equation}\label{ZGFT}
	Z_{\text{GFT}} = \int \mathcal{D}\varphi\; e^{-S[\varphi]} = \sum_h  \lambda^{n_V(h)}\mathcal{A}_h\,,
\end{equation}
where $n_V(h)$ is the number of vertices in $h$ and $\mathcal{A}_h$ are Feynman amplitudes. Remarkably, the sum \eqref{ZGFT} is over graphs that for suitable choices of $V[\varphi]$ can be seen as discrete ``histories of geometry'', and whose Feynman amplitudes $\mathcal{A}_h$ are in correspondence with spin foam amplitudes \cite{Reisenberger_2000,BCmodelfromGFT}. In this sense, the Feynman amplitudes of a GFT with action \eqref{GFTaction} can be associated with a discrete quantum gravity path integral, and the expansion \eqref{ZGFT} generates a sum over two-complexes (or discrete spacetime histories), weighted by the coupling $\lambda$ of the interaction term.

GFT models for a full theory of quantum gravity still remain formal as it is not clear how to make mathematical sense of \eqref{ZGFT}. It is already very difficult to compute individual transition amplitudes $\mathcal{A}_h$ between quantum geometries. Here we will focus on a \textit{canonical quantisation}, which provides very useful insights into the cosmological sector of GFT \cite{Gielen_2016,GFTcosmoLONGpaper}. This quantisation is the simplest in an approximation in which one neglects the interaction $V[\varphi]$, which is what we will do in the following. Restriction to the free theory, while motivated by computational simplicity, is often justified when looking at cosmological models: neglecting correlations between ``quanta of geometry'' can be interpreted as describing GFT configurations of high symmetry, associated with macroscopic homogeneous spacetimes.

A geometrical interpretation of GFT is obtained by associating a 3-simplex (tetrahedron) with the group field $\varphi(g_I,\chi)$. In a dual picture, one can think equivalently of $\varphi(g_I,\chi)$ as an abstract node with four links labelled by $SU(2)$ arguments, and an additional real label $\chi$. This is equivalent to the way in which four-valent spin network nodes represent geometric tetrahedra in LQG.  The nomenclature spin network derives from the Peter--Weyl theorem, which allows decomposing $\varphi$ as
\begin{equation}\label{Peter}
	\begin{aligned}
	\varphi(g_I,\chi)& = \sum_J \varphi_J(\chi) D_J(g_I)\,, \\  	D_J(g_I)& = \sum_{n_I} R_{n_I}^{j_I, \imath} \prod_{a=1}^{4} \sqrt{2j_a+1} D^{(j_a)}_{m_a, n_a}(g_a)\,.
	\end{aligned}
\end{equation}
Here $\varphi_J (\chi)$ are complex functions (subject to reality conditions) and the compact notation for the modes $\pm J = (j_I,\pm m_I,\imath)$ encodes representation (or spin) labels $j_I \in \mathbb{N}_0/2$, magnetic indices $m_I,n_I \in \left[-j_I,j_I\right]$ and intertwiner labels $\imath$. In the convolution $D_J(g_I)$, $R_{n_I}^{j_I, \imath}$ are intertwiners\footnote{$SU(2)$ intertwiners are equivariant linear maps from the tensor product $\bigotimes_I {j_I}$ to the trivial representation. They form a vector space, with basis labelled by $\iota$. Such tensors appear in \eqref{Peter} because of property \eqref{groupfield}.} for the spins $j_I$, and $D^{(j)}_{m, n}(g)$ are Wigner D-matrices for the irreducible unitary representations of $SU(2)$. The mode decomposition \eqref{Peter} shifts the focus from group variables to the  more convenient spin variables. In this representation, the free GFT action (with kinetic term given by \eqref{kinetic}) has the form \cite{relham_Wilson_Ewing_2019,*relhamadd}
\begin{equation}\label{freeGFT}
	S[\varphi] =  \frac{1}{2} \int {\rm d}\chi \sum_{J} \varphi_{-J}(\chi)\left( K^{(0)}_{J}+K^{(2)}_{J}\partial_\chi^2\right) \varphi_J(\chi)\,.
\end{equation}

A canonical quantisation of \eqref{freeGFT} can now be obtained in a \textit{deparametrised} formalism. (There is another approach commonly used in GFT which we call \textit{algebraic quantisation}, see section \ref{algebraicsection}.) In a deparametrised approach we choose a degree of freedom to parametrise the others {before} quantisation; here the obvious candidate is the matter clock $\chi$. One then performs the Legendre transform, introducing a conjugate momentum $\pi_J (\chi)$ to the group field and finding  a  relational Hamiltonian \cite{relham_Wilson_Ewing_2019,*relhamadd}
\begin{equation}\label{classicalH}
	\mathcal{H} = -\frac{1}{2} \sum_{J} \left[\frac{\pi_{J}(\chi)\pi_{-J}(\chi)}{K^{(2)}_{J}}+K^{(0)}_{J}\varphi_{J}(\chi)\varphi_{-J}(\chi)\right] \,.
\end{equation}
The Hamiltonian \eqref{classicalH} defines dynamics of any observable via Poisson brackets or, in the quantum theory defined in the Heisenberg picture, of any operator via the Heisenberg equation. Adopting the Heisenberg picture from now on, we promote the field and its momentum to operators with canonical (equal-time) commutation relations
\begin{equation}
	\left[ \hat{\varphi}_J(\chi), \hat{\pi}_{J'}(\chi) \right] = {\rm i}\,\delta_{JJ'} \,.
\end{equation}
As in any bosonic field theory, one can now define ladder operators with commutation relations
\begin{equation}
	\left[ \hat{a}_J(\chi), \hat{a}^\dagger_{J'}(\chi) \right] = \delta_{JJ'} 
\end{equation}
and construct a Fock space, starting from a vacuum $|0 \rangle$ (interpreted as a ``no-geometry'' state) such that $\hat{a}_J(\chi)|0\rangle = 0$. Excitations created by these ladder operators are interpreted as quanta of geometry: a one-particle state $	|\, \vcenter{\hbox{\adjincludegraphics[width=.025\textwidth]{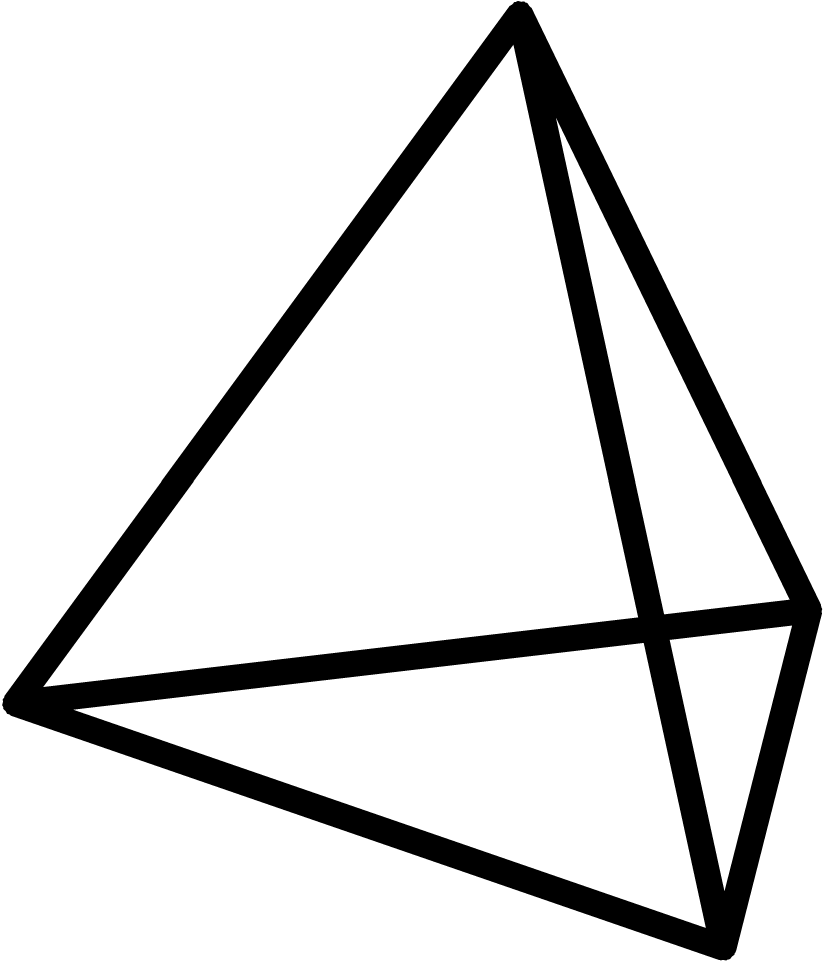} }}\rangle= \hat{a}^\dagger_J(\chi) |0\rangle$ represents a quantum tetrahedron (or four-valent node) decorated with a real variable $\chi$ and group-theoretic information encoded in $J$.

The specific expression of the Hamiltonian \eqref{classicalH} in terms of ladder operators depends on the signs of the kinetic terms $K^{(0)}$ and $K^{(2)}$ in \eqref{freeGFT}: when they have opposite signs, the Hamiltonian is of squeezing type 
\begin{equation}\label{sqH}
	\begin{aligned}
	\hat{\mathcal{H}} &=  \frac{1}{2} \sum_J \omega_{J} \left(\hat{a}^{\dagger}_{J}\hat{a}^{\dagger}_{-J}+\hat{a}_{J}\hat{a}_{-J}\right)\,,\\  \omega_{J} &= -\sgn\left(K^{(0)}_{J}\right)\sqrt{\left|K^{(0)}_{J}/K^{(2)}_{J}\right|}\,.
	\end{aligned}
\end{equation}
For modes for which $K^{(0)}$ and $K^{(2)}$ have the same sign, on the other hand, one obtains the Hamiltonian of a harmonic oscillator. One also defines a number operator in the usual way as
\begin{equation}\label{number}
	\hat{N}(\chi) = \sum_J \hat{a}^{\dagger}_J(\chi)\hat{a}_J(\chi) = \sum_J \hat{N}_J(\chi)\,.
\end{equation}
The squeezing Hamiltonian \eqref{sqH} creates pairs of GFT quanta (with opposite magnetic indices) from the vacuum state, providing a compelling picture for an expanding cosmological geometry. The number of quanta in these unstable modes will then (for generic initial conditions) quickly exceed the number of quanta in any of the stable modes, for which the particle number is constant.

To obtain a cosmological interpretation, a central role is played by the volume operator 
\begin{equation}\label{volume}
	\hat{V}(\chi) = \sum_J v_{J} \hat{a}^{\dagger}_J(\chi)\hat{a}_J(\chi) = \sum_J \hat{V}_J(\chi)\,,
\end{equation}
where the $v_J$'s correspond to the volumes of quanta with representation data $J$. These volume values can be formally obtained from a geometrical quantisation of tetrahedra in terms of $SU(2)$ recoupling theory \cite{Barbieri_1997,*Baez_1999,*PolyhedraIta}, and they have been thoroughly described in the LQG literature (see, e.g., \cite{Bianchi_Haggard_vol,*Bianchi_Haggard_letter}) where they define the volume operator acting on four-valent spin network nodes. The operator \eqref{volume} describes a \textit{global} notion of spatial volume seen as the sum of many discrete building blocks carrying their own (quantum) volume. 

\subsection{$\mathfrak{su(1,1)}$ algebra and FLRW cosmology}\label{Subsu11FLRW}

In addition to restricting to the free GFT, one often also restricts the setup to a single Peter--Weyl mode $J$, or a coupled pair of modes $\{J,-J\}$. In the latter case, one can choose ``symmetric'' initial conditions $\hat{a}_J (0) = \hat{a}_{-J}(0)$ which are preserved under time evolution \cite{relham_Wilson_Ewing_2019,*relhamadd}. The Hamiltonian then effectively describes single-mode squeezing. A more direct way of picking out a single mode is to consider a $J$ with magnetic indices $m_I = 0$ (so that $J=-J$); this is a somewhat mild assumption as no geometrical observable depends on the values of the magnetic indices. It turns out that considering only a single mode (i.e., only excitations of the same ``type'') is enough to obtain the correct cosmological dynamics of a flat FLRW universe \cite{Oriti_2016,*BOriti_2017}. Moreover, for a wide class of models such an assumption can be justified by the fact that some modes (those for which $|\omega_J|$ in \eqref{sqH} is largest) grow faster than all others and so eventually dominate \cite{Gielen_lowspin}. This means that an effective restriction to a single mode (or a small number of physically indistinguishable modes) would also emerge dynamically at sufficiently late times, which is the regime we might want to compare with classical cosmology. We will use this restriction throughout the rest of the paper, and only study quantum states for a single Peter--Weyl mode.

Restricting to a single mode means that the sums over modes $J$ trivialise to only one term. In order to simplify the notation, we will henceforth drop the index $J$ in our single-mode expressions. We now deal with a quantum system described by bosonic operators $\hat{a}(\chi)$ and $\hat{a}^\dagger(\chi)$ with $\left[ \hat{a}(\chi), \hat{a}^\dagger(\chi) \right] = 1$. The main operators of interest for cosmological purposes are the Hamiltonian \eqref{sqH} and the volume \eqref{volume}, which reduce to
\begin{equation}\label{HV}
	\begin{aligned}
	\hat{{H}} &=  -\frac{\omega}{2}  \left(\hat{a}^{\dagger 2}+\hat{a}^2\right)	\,,\\   \hat{{V}}(\chi) &= v \, \hat{{N}}(\chi) = v\, \hat{a}^{\dagger}\hat{a}  \,,
	\end{aligned}
\end{equation}
where $v$ can be thought of as the volume of one GFT quantum. {As mentioned}, the operators \eqref{HV} generate the Lie algebra $\mathfrak{su(1,1)}$ extended by a central element \cite{Gielen_2020,Bojo_2019}. The algebra is closed by adding 
\begin{equation}\label{ccc}
	\hat{C} = {\rm i} \frac{v}{2} \left(\hat{a}^{\dagger 2}-\hat{a}^2\right)\,,
\end{equation}
which would be related to the ``Thiemann complexifier'' in some analogous LQG models \cite{CVHalgebra}. Here \eqref{ccc} does not have a direct physical interpretation, but determines whether the resulting cosmology has a time-reversal symmetry. The $\mathfrak{su(1,1)}$ algebra of these operators follows from their composition in terms of ladder operators: one traditionally defines the three possible quadratic combinations $\hat{K}_0 = \frac{1}{4}(\hat{a}^\dagger\hat{a}+\hat{a}\hat{a}^\dagger)$, $\hat{K}_+ = \frac{1}{2} \hat{a}^{\dagger 2}$ and $\hat{K}_- = \frac{1}{2} \hat{a}^2$, which satisfy the $\mathfrak{su(1,1)}$ algebra
\begin{equation}\label{su11}
	\left[\hat{K}_0,\hat{K}_\pm \right] = \pm \hat{K}_\pm \,, \qquad \left[\hat{K}_- , \hat{K}_+ \right] =2\hat{K}_0 \,.
\end{equation}
In our case, the GFT operators relate to these $\mathfrak{su(1,1)}$ generators as
\begin{align}
%	\begin{aligned}
	\hat{H} &= -\omega (\hat{K}_++\hat{K}_-) \,, \nonumber\\  \hat{V} &= 2 v \hat{K}_0 - \frac{v}{2} \,, \\ \hat{C} &= {\rm i} v (\hat{K}_+-\hat{K}_-) \,,\nonumber
%	\end{aligned}
\end{align}
and the algebra closes as
\begin{align}\label{algebra}
	\left[ \hat{V}, \hat{H} \right] &= 2 {\rm i} \omega \hat{C} \,, \nonumber \\  \left[\hat{C},\hat{V}\right] &=2{\rm i} \frac{v^2}{\omega} \hat{H} \,, \\  \left[\hat{C}, \hat{H}\right] &= 2 {\rm i} \omega \left(\frac{v}{2} + \hat{V}\right) \,, \nonumber
\end{align}
where we see the central element (identity operator) appearing in the third relation.

{We stress} that while we focus on GFT in this paper, the $\mathfrak{su(1,1)}$ structure (together with a cosmological interpretation) is the only necessary ingredient for all our main results. Relations analogous to \eqref{algebra} are described in other realisations of $\mathfrak{su(1,1)}$ quantum cosmology \cite{Bojo_2019, HarC1, *HarC2}, for example in loop quantum cosmology (where this algebra commonly appears \cite{su11LQC,Boden1, *Boden2, *Boden3}).

We can now turn to the dynamics of such operators. $\hat{H}$ determines the evolution of any other operator $\mathcal{\hat{O}}$ via the Heisenberg equation
\begin{equation}
	{\rm i}\frac{{\rm d} \hat{\mathcal{O}}}{{\rm d}\chi} = [\hat{\mathcal{O}},\hat{{H}}] \,.
\end{equation}
From this, one can obtain the solutions \cite{Gielen_2020} (here and in the following we will adopt a somewhat unusual notation $f_\alpha$ instead of $f(\alpha)$ for trigonometric and hyperbolic functions; this is to save space in lengthy expressions below)
\begin{align}\label{vofchi}
	\hat{V}(\chi) &= - \frac{v}{2}+\left( \hat{V} + \frac{v}{2}\right)\cosh_{2 \omega \chi} + \hat{C} \sinh_{2 \omega \chi}  \, ,
	\\\label{cofchi}
	\hat{C}(\chi) &=\hat{C} \cosh_{2 \omega \chi} + \left( \hat{V} + \frac{v}{2}\right)\sinh_{2 \omega \chi}    \, ,
\end{align}
where $\hat{V} = \hat{V}(0)$ and $\hat{C} = \hat{C}(0)$. We can now see that, while $\hat{V}$ represents the volume at $\chi=0$, the presence of $\hat{C}$ determines whether the volume evolution \eqref{vofchi} has a symmetry under $\chi\rightarrow-\chi$. Since we are working in the Heisenberg picture, the solutions \eqref{vofchi} and \eqref{cofchi} do not refer to any choice of quantum state; in fact, \eqref{vofchi} is all one needs to obtain an effective Friedmann equation. Taking expectation values one finds
\begin{widetext}
\begin{equation}\label{Friedmann}
	\left(\frac{1}{\langle \hat{V}(\chi)\rangle } \frac{{\rm d} \langle \hat{V}(\chi)\rangle}{{\rm d} \chi}\right)^2 = 4\omega^2 \left( 1+\frac{v}{\langle \hat{V}(\chi)\rangle} -\frac{1}{\langle \hat{V}(\chi)\rangle^2}\left[\langle \hat{V}\rangle^2 +v \langle \hat{V}\rangle - \langle \hat{C}\rangle^2 \right]\right)\,.
\end{equation} 
\end{widetext}
For large volumes (or late times $\chi \rightarrow \pm \infty$), \eqref{Friedmann} is consistent with the classical Friedmann equation\footnote{\label{classicalFE}In general relativity, the Friedmann equation for the spatial volume $V$ of a flat FLRW cosmology with a massless scalar field $\chi$, using a gauge where $\chi$ is the time coordinate, can be written as $\left(\frac{1}{V}\frac{\dd V}{\dd\chi}\right)^2 = 12\pi G $.	} provided the identification between the GFT coupling and Newton's constant $\omega^2 = 3\pi G$. The two subleading contributions can be seen as GFT corrections to classical cosmology. In particular, the $1/\langle \hat{V}(\chi)\rangle^2$ term is responsible for the generic resolution of the Big Bang singularity, which is replaced with a cosmological bounce through a minimal nonsingular volume.

We point out that \eqref{Friedmann} holds regardless of whether the quantum state one uses to compute expectation values is a pure or mixed state. All that is needed to obtain \eqref{Friedmann} is a (linear) operation mapping operators to their expectation values, and the density matrix expression $\langle \hat{V}(\chi)\rangle = \tr\left( \hat{\rho} \, \hat{V}(\chi)  \right)$ is as good as the pure-state evaluation $\langle \hat{V}(\chi)\rangle=\langle\psi|\hat{V}(\chi)|\psi\rangle$. This point was not stressed in \cite{Gielen_2020} where \eqref{Friedmann} was obtained, and will allow us to investigate semiclassical properties of mixed (in particular \textit{thermal}) states in later sections. As we review below, only a few types of states (mainly Fock coherent states) are typically used in GFT cosmology. This is where we wish to expand the literature: we will discuss criteria that can classify states as semiclassical and present Gaussian states as the most general family of semiclassical states for our theory.

\section{Semiclassical properties and candidate states}\label{StatesSection}
%----------------------------------------------------------------------------

While true for any quantum state, the effective Friedmann equation \eqref{Friedmann} is a relation between expectation values only. To claim that this equation is a good description of the dynamics of cosmological observables, one needs to adopt quantum states that show some semiclassical features, such as coherent states. More generally, one needs to specify criteria for any candidate state for cosmology to be considered as semiclassical. Here we focus on two criteria that are commonly used: the study of relative uncertainties and the Robertson--Schr\"odinger uncertainty principle. We define variances and covariances for any operators $\hat{A}$ and $\hat{B}$ as
\begin{align}
	(\Delta \hat{A})^2 &= \langle \hat{A}^2\rangle - \langle \hat{A} \rangle^2 \,, \label{variance}\\ 
	\Delta (\hat{A} \hat{B}) &= \frac{1}{2}\langle\{\hat{A},\hat{B}\}\rangle-\langle \hat{A} \rangle \langle \hat{B} \rangle \,, \label{covariance}
\end{align}
where $\{ \cdot,\cdot\}$ is the anticommutator. 

Our first criterion for semiclassicality would be to require that relative uncertainties $(\Delta \hat{A})^2/\langle \hat{A}\rangle^2$ be small at least in a large-volume or late-time regime where the classical theory is expected to emerge; here $\hat{A}$ could be either the Hamiltonian $\hat{H}$ or volume $\hat{V}$. One can also check what happens to the operator $\hat{C}$ defined in \eqref{ccc}, even though its interpretation is less transparent, hence it is unclear whether this operator would need to be semiclassical.

There is another characterisation of semiclassical states that makes use of the quantities \eqref{variance} and \eqref{covariance}, namely the saturation of the Robertson–Schr\"odinger (RS) uncertainty principle \cite{RS,*RS2}. For the GFT operators \eqref{HV}, the uncertainty principle reads
\begin{equation}\label{RSprinciple}
	(\Delta \hat{V})^2(\Delta \hat{H})^2 \geq |\Delta (\hat{V}\hat{H})|^2 +\omega^2\langle\hat{C}\rangle^2 \,.
\end{equation}
For basic examples in standard quantum mechanics an inequality of this type is saturated (it becomes an equality) for canonically conjugate pairs when using coherent (or more generally Gaussian) states; but in general it is not guaranteed that there are states for which \eqref{RSprinciple} can be minimised. As the volume evolves in time, the RS uncertainty principle \eqref{RSprinciple} is a statement for each $\chi$. 

In the context of GFT cosmology or quantum cosmology in general, demanding small relative uncertainties seems physically more relevant than minimising uncertainties by demanding equality in \eqref{RSprinciple}; nothing in \eqref{RSprinciple} requires both sides to be small in any sense, whereas the Universe appears to be sharp to observations, without quantum effects on large scales. Hence, we would say that a good candidate state for GFT cosmology models primarily needs to show small relative uncertainties. As we will see shortly, (Fock) coherent states have this property; we will also define more general states that are semiclassical in this sense.

From \eqref{vofchi} we can derive the $\chi$-dependent form of the volume variance as well as that of the covariance between the volume and the Hamiltonian,
\begin{align}
	(\Delta \hat{V}_\chi )^2 &= (\Delta \hat{V} )^2 \cosh^2_{2\omega \chi} +(\Delta \hat{C} )^2 \sinh^2_{2\omega \chi}  \label{chivar} \\ & \quad+ \Delta (\hat{V}\hat{C})\sinh_{4\omega \chi} \,,\nonumber\\ 
	\Delta (\hat{V}_\chi\hat{H}) &= \Delta (\hat{V} \hat{H}) \cosh_{2\omega \chi} + \Delta (\hat{C}\hat{H}) \sinh_{2\omega \chi}\,,\label{chicovar}
\end{align}
where from now on we use subscripts to indicate time-dependent operators; operators with no subscript refer to initial conditions (i.e., to $\chi=0$). We can then immediately derive the large-volume limit of relative uncertainties by taking $\chi \rightarrow \pm \infty$ in these expressions: ${(\Delta \hat{H} )^2}/{\langle \hat{H}\rangle^2}$ does not evolve in time, but for the relative volume fluctuations we find using \eqref{chivar} and \eqref{vofchi}
\begin{equation}\label{rs1}
	\frac{(\Delta \hat{V}_\chi )^2}{\langle \hat{V}_\chi \rangle^2} \quad\overset{\chi\rightarrow\pm\infty}{\longrightarrow}\quad \frac{(\Delta \hat{V} )^2+(\Delta \hat{C} )^2\pm2\Delta (\hat{V}\hat{C})}{\left(\langle \hat{V} \rangle + \frac{v}{2} \pm\langle \hat{C}\rangle \right)^2}\,.
\end{equation}
To verify the RS uncertainty principle, we would also need the limit 
\begin{equation}\label{rs2}
	\begin{aligned}
&	\frac{(\Delta (\hat{V}_\chi\hat{H}))^2}{\langle \hat{V}_\chi \rangle^2 \langle \hat{H} \rangle^2}+	\omega^2 \frac{\langle \hat{C}_\chi \rangle^2}{\langle V_\chi \rangle^2 \langle \hat{H} \rangle^2}\\ & \quad\overset{\chi\rightarrow\pm\infty}{\longrightarrow}\quad
\frac{\left( \Delta (\hat{V} \hat{H}) \pm  \Delta (\hat{C} \hat{H})\right)^2}{\left(\langle \hat{V} \rangle + \frac{v}{2} \pm\langle \hat{C}\rangle \right)^2 \langle \hat{H} \rangle^2} + \frac{\omega^2}{\langle \hat{H} \rangle^2} \,.
	\end{aligned}
\end{equation}
All these quantities are determined by the initial conditions only. Notice that the late-time limit $\chi\rightarrow +\infty$ in general differs from the limit $\chi\rightarrow -\infty$ (as indicated by the $\pm$ notation) so that the asymmetry described by the quantity $\hat{C}$ is manifest here.

One might also be interested in the evolution of these quantities beyond the strict infinite-volume limit $\chi\rightarrow\pm\infty$. We will show some examples for general evolution of the RS inequality \eqref{RSprinciple} and the relative uncertainties of the volume operator.  For analytical results, we derive expansions in powers of the inverse volume as 
\begin{equation}\label{expansion}
	\frac{(\Delta \hat{V}_\chi)^2}{\langle \hat{V}_\chi \rangle^2} =  \mathcal{A} + \mathcal{B} \frac{v }{\langle \hat{V}_\chi \rangle}+ \mathcal{C} \frac{v^2}{\langle \hat{V}_\chi \rangle^2} + \cdots \,,
\end{equation} 
where $\mathcal{A}$, $\mathcal{B}$, $\mathcal{C}$, $\dots$ are functions of initial conditions. Such an expansion captures very well the full evolution as soon as we are in the macroscopic regime $\langle\hat{V}_\chi\rangle\gg v$.

We shall see that the saturation of \eqref{RSprinciple} does not necessarily indicate that the states under question have small relative fluctuations; conversely, states such as the simplest Fock coherent states, which are semiclassical by looking at relative fluctuations, fail to minimise \eqref{RSprinciple}.  In the rest of this section we briefly review quantum states that have been investigated in the context of GFT cosmology. We explicitly check whether the RS uncertainty principle is minimised and we obtain the exact dynamics of the relative uncertainties for the volume in closed form. These properties will serve as comparison for the new family of states presented in section \ref{GaussSection}.

\subsection{Coherent states}

The most commonly used states in the GFT cosmology literature are Fock coherent states, introduced already in \cite{GFTcosmoLONGpaper} and used in different ways in the deparametrised formalism \cite{relham_Wilson_Ewing_2019,*relhamadd} and the algebraic approach \cite{Oriti_2016,*BOriti_2017,Marchetti2021}. As is customary in bosonic theories, coherent states can be defined via the action of the displacement operator $\hat{D}(\alpha)$ on the Fock vacuum as
\begin{equation}\label{CS}
	\begin{aligned}
	|\alpha\rangle &= \hat{D}(\alpha) |0\rangle\,,  \\  \hat{D}(\alpha)  &= e^{\alpha \hat{a}^\dagger -\bar{\alpha}\hat{a}}\,,  \qquad \alpha \in \mathbb{C}\,.
	\end{aligned}
\end{equation}
These have the key property that, at $\chi=0$, $\hat{a}(0) |\alpha \rangle = \alpha |\alpha\rangle$. Appendix \ref{UncertApp} contains a table with all the quantities of interest computed with the state \eqref{CS}.

One can easily check that the RS uncertainty principle \eqref{RSprinciple} for $\hat{V}$ and $\hat{H}$ is \textit{not} saturated by coherent states. For instance (using table \ref{TabCohSqu} in appendix \ref{UncertApp}), we see that at $\chi=0$ \eqref{RSprinciple} reads
\begin{equation}\label{coherentUR}
	\frac{v^2\omega^2}{2}|\alpha|^2 +v^2\omega^2  |\alpha|^4 \geq v^2\omega^2|\alpha|^4\,.
\end{equation}
This feature occurs because \eqref{HV} and \eqref{ccc} are $\mathfrak{su(1,1)}$ compositions of the bosonic ladder operators, whereas the Fock coherent state is coherent with respect to $\hat{a}$ and $\hat{a}^\dagger$. One can in fact show that \eqref{RSprinciple} is \textit{never saturated} by coherent states. We refer to appendix \ref{UncertApp} where we report explicitly the analytical expressions representing the general case of \eqref{RSprinciple}; such a minimisation does not happen at any time and in particular not as $\chi\rightarrow\pm \infty$, where the system is meant to become semiclassical.

As one might expect, this does not really spoil the semiclassical nature of coherent states in the sense of relative uncertainties. Decomposing $\alpha$ into modulus and argument as $\alpha = |\alpha| \exp ({\rm i} \vartheta)$, one can see that the relative uncertainties at $\chi=0$ (again, see appendix \ref{UncertApp}),
\begin{equation}\label{RUCS0}
	\begin{aligned}
	\frac{(\Delta \hat{V})^2_\text{C}}{\langle \hat{V} \rangle_{\text{C}}^2} &=  \frac{1}{|\alpha|^2} \,,\\
\frac{(\Delta \hat{H})^2_{\text{C}}}{\langle \hat{H} \rangle_{\text{C}}^2} &=  \frac{4 |\alpha|^2+2}{4|\alpha|^4 \cos_{2\vartheta}^2} \,,\\
\frac{(\Delta \hat{C})^2_{\text{C}}}{\langle \hat{C} \rangle_{\text{C}}^2} &=    \frac{4 |\alpha|^2   +2}{4|\alpha|^4 \sin_{2\vartheta}^2} \,,
	\end{aligned}
\end{equation}
can be made arbitrarily small by choosing appropriate $|\alpha|$ and avoiding parameters for which $\vartheta$ is a multiple of $\frac{\pi}{4}$ (or of the form $\frac{\pi}{4}+k\frac{\pi}{2}$ with $k\in\mathbb{Z}$, if we are only interested in small $(\Delta \hat{H})^2_{\text{C}}/\langle \hat{H} \rangle_{\text{C}}^2$). Away from $\chi=0$, from \eqref{vofchi} and \eqref{chivar} one finds the relative volume relative uncertainty
\begin{equation}\label{RUCS}
	\frac{(\Delta \hat{V}_\chi)^2_{\text{C}}}{\langle \hat{V}_\chi \rangle_{\text{C}}^2} = \frac{\left(4 | \alpha|^2+1\right) \cosh_{4 \omega  \chi }+ 4|\alpha|^2 \sin_{2\vartheta} \sinh_{4 \omega  \chi} -1}{\left(\left(2 | \alpha|^2+1\right)\cosh_{2 \omega  \chi } +2|\alpha|^2 \sin_{2\vartheta} \sinh_{2 \omega  \chi}-1 \right)^2}\,,
\end{equation}
By choosing $|\alpha|$  to be large, this can be made arbitrarily small at all times: consider the asymptotic behaviour of \eqref{RUCS} for large $|\alpha|$,
\begin{equation}\label{largealphaCS}
	\frac{(\Delta \hat{V}_\chi)^2_{\text{C}}}{\langle \hat{V}_\chi \rangle_{\text{C}}^2} \quad \sim \quad \frac{1}{|\alpha|^2}\frac{\sin_{2 \vartheta } \sinh_{4   \omega\chi} +\cosh_{4  \omega\chi }}{ (\sin_{2 \vartheta} \sinh_{2   \omega\chi }+\cosh_{2   \omega\chi} )^2} \,,
\end{equation}
and notice that $1/|\alpha|^2$ multiplies a bounded function in $\chi$. For late times, we recover the results of \cite{Gielen_2020}
\begin{equation}\label{ACS}
	\frac{(\Delta \hat{V}_\chi)^2_{\text{C}}}{\langle \hat{V}_\chi \rangle_{\text{C}}^2} \quad \overset{\chi\rightarrow \pm \infty}{\longrightarrow}  \quad \frac{2(1 +4|\alpha|^2(1\pm \sin_{2\vartheta}))}{(1+2|\alpha|^2(1\pm\sin_{2\vartheta}))^2} =:	\mathcal{A}_{\text{C}}\,.
\end{equation}
Again, this becomes arbitrarily small for large $|\alpha|$ and avoiding the values $\vartheta=\frac{\pi}{4}+k\frac{\pi}{2}$.

We can also expand \eqref{RUCS} in inverse volume powers, finding
\begin{equation}\label{secondorderCS}
	\frac{(\Delta \hat{V}_\chi)^2_{\text{C}}}{\langle \hat{V}_\chi \rangle_{\text{C}}^2} =  \mathcal{A}_{\text{C}} \left(1+ \frac{v}{\langle \hat{V}_\chi \rangle}\right) +  \mathcal{C}_{\text{C}}\frac{v^2}{\langle \hat{V}_\chi \rangle^2} + \mathcal{O}\left(\frac{1}{\langle \hat{V}_\chi \rangle^4}\right) \,,
\end{equation}
with
\begin{equation}\label{CCS}
	\mathcal{C}_{\text{C}} = -\frac{| \alpha | ^2 \pm 2 | \alpha | ^4 (\sin_{2 \vartheta} \pm1) \left(| \alpha | ^2 (\cos_{4 \vartheta }+1)+3\right)}{\left(2 | \alpha | ^2 (\sin_{2 \vartheta}  \pm1)\pm1\right)^2} \,.
\end{equation}
We see that the $1/\langle \hat{V}(\chi) \rangle$ correction is such that in \eqref{expansion} $\mathcal{B}_{\text{C}}= \mathcal{A}_{\text{C}}$, which is similar to the terms in the Friedmann equation \eqref{Friedmann}. Higher contributions can be found, but they only minimally improve the expansion, whose key behaviour is already captured at the $1/\langle \hat{V}(\chi) \rangle^2$ order. 

\subsection{Squeezed states}\label{SSSection}

Mimicking standard quantum mechanics notation, squeezed states can be defined via the action of the squeezing operator $\hat{S}(z)$ on the Fock vacuum as
\begin{equation}\label{SS}
	\begin{aligned}
			|z\rangle &= \hat{S}(z) |0\rangle \,, \\ \hat{S}(z) &= e^{\frac{1}{2}\left(z\hat{a}^{\dagger 2} -\bar{z}\hat{a}^2\right)} \,, \qquad z\in \mathbb{C}\,.
	\end{aligned}
\end{equation}
We decompose $z$ as $z=r e^{{\rm i}\psi}$, where $r$ and $\psi$ are real parameters.

These squeezed states can be seen as part of the Perelomov--Gilmore class of coherent states \cite{Perelomov,*Gilmore} associated with $SU(1,1)$; this is how they were introduced for GFT in \cite{Gielen_2020}. As described in \cite{Gielen_2020}, the volume operator \eqref{HV} is bounded from below only in the $\mathfrak{su(1,1)}$ representations of the positive ascending series; when one restricts to the cases of interest for GFT,\footnote{The representations of the positive discrete series are labelled by a real parameter $k$ called Bargmann index. Using the bosonic realisation of $\mathfrak{su(1,1)}$ \eqref{su11} and wishing to include the Fock vacuum among the eigenstates of the volume operator, one is led to choose $k=1/4$, for which all the results of \cite{Gielen_2020} coincide with the ones described here.} the Perelomov--Gilmore coherent states coincide exactly with the squeezed states that we define here.

Contrary to coherent states, one can readily find that squeezed states \textit{do} saturate the RS uncertainty principle \eqref{RSprinciple} for the operators $\hat{V}$ and $\hat{H}$. Using again table \ref{TabCohSqu} in appendix \ref{UncertApp}, at $\chi=0$ one explicitly has
\begin{equation}\label{squeezeUR}
	\begin{aligned}
&	\frac{v^2 \omega ^2}{16} \sinh ^2_{2 r} \left(2 \sinh ^2_{2 r} \cos_{2 \psi} +\cosh_{4 r}+3\right) \\ &= \frac{v^2 \omega ^2}{16}  \cos ^2_\psi \sinh ^2_{4 r} + \frac{v^2\omega^2}{4} \sin ^2_\psi \sinh ^2 _{2r}  \,.
	\end{aligned}
\end{equation}
This minimisation happens because we are interested in uncertainties of the GFT operators \eqref{HV} and \eqref{ccc}, which form the $\mathfrak{su(1,1)}$ structure that squeezed states are built on. ({An analogous} result for $SU(1,1)$ coherent states in loop quantum cosmology is reported in \cite{su11LQC}.) Turning on time dependence, we find that the uncertainty principle is indeed an exact equality throughout the whole evolution for the state in \eqref{SS}. Again we refer to appendix \ref{UncertApp} for the analytical expressions at generic times; there we show that the RS uncertainty principle is minimised for all values of $\chi$ and in particular in the late-time limit $\chi\rightarrow\pm \infty$.

The minimisation of the RS principle does  not necessarily mean that relative uncertainties of cosmological observables are small, and indeed we find at $\chi=0$ (see table \ref{TabCohSqu} in appendix \ref{UncertApp})
\begin{equation}\label{RUSS0}
	\begin{aligned}
	\frac{(\Delta \hat{V})^2_{\text{S}}}{\langle \hat{V} \rangle_{\text{S}}^2} &= 2\coth ^2_r   \,,\\
\frac{(\Delta \hat{H})^2_{\text{S}}}{\langle \hat{H} \rangle_{\text{S}}^2} &=  2+2 \sec ^2_\psi  \text{csch}^2_{2 r}\,,\\
\frac{(\Delta \hat{C})^2_{\text{S}}}{\langle \hat{C} \rangle_{\text{S}}^2} &=  2 + 2 \csc ^2_\psi  \text{csch}^2_{2 r} \,.
	\end{aligned}
\end{equation}   
All these quantities are bounded from below by 2. One can still check whether the situation improves with time evolution; a minimal requirement for semiclassicality is that relative uncertainties are only small at large volume. Using \eqref{vofchi} and \eqref{chivar} one can readily write down the exact time evolution of the relative uncertainties as

	\begin{align}\label{RUSS}
	\frac{(\Delta \hat{V}_\chi)^2_{\text{S}}}{\langle \hat{V}_\chi \rangle_{\text{S}}^2} &= 2 \, \frac{ \sin_\psi  \sinh_{2 r} \sinh_{2 \omega  \chi} +\cosh_{2 r} \cosh _{2 \omega  \chi}+1}{\sin_\psi  \sinh_{2 r} \sinh_{2 \omega  \chi} +\cosh_{2 r} \cosh_{2 \omega  \chi} -1} \nonumber \\& = 2\left(1 + \frac{v}{\langle \hat{V}_\chi\rangle}\right)\,.
	\end{align}

Hence, the lower bound of 2 for the relative uncertainty holds at all times; a uniform large-volume limit of 2 was already found in \cite{Gielen_2020}. 

As a final remark on squeezed states, we point out that a ``dipole condensate'' state
\begin{equation}\label{dipoleDP}
	| \xi \rangle = \exp \left(\frac{1}{2} \xi \, \hat{a}^\dagger \hat{a}^\dagger\right) |0\rangle \,, \quad \qquad \xi\in \mathbb{C}\,,
\end{equation}
is nothing else but a non-normalised squeezed state. States similar to \eqref{dipoleDP} were introduced as possible condensate-like states in the early stages of GFT cosmology \cite{GFTcosmoLONGpaper} (we will return to a discussion of these states in the algebraic approach later).
Given the norm 
\begin{equation}
	\langle \xi | \xi \rangle = \frac{1}{\sqrt{1-|\xi|^2}} \,,
\end{equation}
we should assume $|\xi|<1$ in order to obtain a normalisable state.

To see that \eqref{dipoleDP} is a squeezed state, we write the squeezing operator $\hat{S}(z)$ in ``normal form'' \cite{SqueezeNORMAL}
\begin{equation}
	\begin{aligned}
	\hat{S}(z) = &\exp\left(  \frac{z}{2 |z|} \tanh |z| \; \hat{a}^{\dagger 2}  \right)\\ &\times \exp\left(- \ln \cosh |z| \left(\hat{a}^\dagger \hat{a} + \frac{1}{2}\right) \right)\\ &\times \exp\left(  - \frac{\bar{z}}{2 |z|} \tanh |z| \; \hat{a}^{2} \right) \,,
	\end{aligned}
\end{equation}
so that one can write a squeezed state  as
\begin{equation}\label{thus}
	| z \rangle = \hat{S}(z) |0\rangle = \frac{1}{\sqrt{\cosh |z|}}  \exp\left( \frac{z}{2 |z|} \tanh |z| \; \hat{a}^{\dagger 2}  \right) |0\rangle \,.
\end{equation}
\eqref{thus} shows that a dipole state \eqref{dipoleDP} is a (rescaled) squeezed state \eqref{SS}, $	| \xi \rangle   = \sqrt{\cosh |z|} \, |z \rangle$, where the dipole parameter $\xi$ and the squeezing parameter $z$ are related by
\begin{equation}\label{map}
	\xi = \frac{z}{|z|}  \tanh |z| \,.
\end{equation}
Since they are just squeezed states, dipole condensates have no chance of being semiclassical according to the criterion of small relative uncertainties.

\section{Gaussian states}\label{GaussSection}

Gaussian states can be defined in several equivalent ways. Traditionally, they are presented in quantum mechanics textbooks as states whose characteristic functions and quasi-probability distributions (also known as Wigner functions) are Gaussian functions. Equivalently, especially in the quantum optics and quantum information literature, Gaussian states are often described as states which are fully determined by the first and second canonical moments only \cite{TextSerafini}. Other characterisations are possible, both physical (as minimum uncertainty states) and mathematical (see \cite{Lucas} for connections to complex structures and symplectic forms).

We will focus on an equivalent but more operational definition of Gaussian states, given as Gibbs states of generic second-order Hamiltonians of bosonic fields \cite{TextSerafini}. Specifically, they can be defined as arising from the action of the displacement operator \eqref{CS} and squeezing operator \eqref{SS} on a thermal state \cite{GaussWigner,*SqueezedAndThermal} (see appendix \ref{GaussThermoApp}),
\begin{equation}\label{GS}
	\hat{\rho}_\text{G} (\alpha, z, \beta)=  \hat{D}(\alpha) \hat{S}(z) \hat{\rho}_{\beta} \hat{S}^\dagger(z) \hat{D}^\dagger(\alpha) \,,
\end{equation} 
where, denoting the usual Fock states by $|n\rangle=(n!)^{-1/2}(\hat{a}^\dagger)^n|0\rangle$,
\begin{equation}\label{thermalstate}
	\hat{\rho}_\beta = \frac{e^{- \beta \hat{a}^\dagger \hat{a}}}{\tr (e^{-\beta \hat{a}^\dagger \hat{a} })}  = (1-e^{-\beta}) \sum_n  e^{-\beta n} |n \rangle \langle n |\,.
\end{equation}
$\beta>0$ is a free parameter, the analogue of the inverse temperature in the usual canonical ensemble.

A key property of Gaussian states is that (in the Schr\"odinger picture) they retain their Gaussian nature under time evolution;  $\hat{U} \hat{\rho}_\text{G} \hat{U}^\dagger $ is also a Gaussian state if $\hat{U}$ is the unitary time evolution operator.\footnote{For a quadratic Hamiltonian the evolution operator can always be decomposed as $	\hat{U} =  e^{{\rm i} \gamma } \hat{S}({z}) \hat{D}({\alpha}) \hat{R}({\phi})$ where $\hat{R}(\phi) = \exp({\rm i} \phi \,\hat{a}^\dagger \hat{a})$ is the rotation operator and $\exp ({\rm i}\gamma)$ a phase factor \cite{Schumaker,*MaRhodes,*MaRhodesMulti}. While a rotation operator can in principle enter the definition of Gaussian states \eqref{GS}, it does not affect any result (see appendix \ref{GaussThermoApp} for details).} This property motivates studies of ``Gaussian'' quantum mechanics, in which one restricts to Gaussian-preserving measurements and transformations and where quadratic Hamiltonians are fundamental \cite{Subtheory1,*Subtheory2}. In this setting one avoids the difficulties that come with higher-order dynamics.

Of course, the family of \textit{pure} Gaussian states is a subset of \eqref{GS} obtained in the vanishing ``temperature'' limit, $\hat{\rho}_{G} \overset{\beta\rightarrow \infty}{\longrightarrow}  \hat{D}(\alpha) \hat{S}(z)|0\rangle \langle 0|\hat{S}^\dagger(z) \hat{D}^\dagger(\alpha) =: |\alpha, z\rangle \langle \alpha,z |$. These states are the well-known displaced squeezed states, which relate nicely to the simpler states discussed in the previous section.

The general class of states \eqref{GS} can straightforwardly be imported in our GFT framework ({and} analogue $\mathfrak{su(1,1)}$ cosmologies) since, as detailed in section \ref{Subsu11FLRW}, we deal with a bosonic system governed by a second-order Hamiltonian \eqref{HV}. The state \eqref{GS} can in fact also be understood along the lines of \cite{Isha_thermalGFT,*Isha_thermalGFT2,*IshaDaniele}, where GFT states are defined as statistical equilibrium states of exponential form $e^{-\hat{\mathcal{O}}}$ for some operator $\hat{\mathcal{O}}$. The parameter $\beta$ in \eqref{thermalstate} is to be taken formally (for instance as the periodicity in the 1-parameter flow of a KMS state or as a Lagrange multiplier \cite{Isha_thermalGFT,*Isha_thermalGFT2,*IshaDaniele}), and does not necessarily relate to a physical notion of temperature. Effectively, given that $\hat{N} =  \hat{a}^\dagger\hat{a}$ represents the number of quanta, $\beta$ in \eqref{thermalstate} could be seen more akin to a chemical potential of a grandcanonical ensemble.

Equipped with the new and generalised family of states \eqref{GS}, we can now turn to the calculation of quantities of interest for harmonic cosmology; appendix \ref{GaussThermoApp} outlines helpful tools for using \eqref{GS} to obtain the following results. First, we compute the expectation value of the three main operators for our models, given in \eqref{HV} and \eqref{ccc}. One finds
\begin{equation}\label{EV}
	\begin{aligned}
		\langle \hat{V} \rangle_\text{G} &=	v \left(| \alpha | ^2 + N_\beta \cosh_{2 r}+\sinh ^2_r\right)  
		\,,\\
		\langle \hat{H} \rangle_\text{G}&=-\frac{\omega}{2}   \left(2|\alpha|^2 \cos_{2\vartheta}+(2 N_\beta+1) \sinh_{2 r} \cos_\psi \right)\,,\\
		\langle \hat{C} \rangle_\text{G} &= \frac{v}{2}  \left( 2|\alpha|^2 \sin_{2\vartheta}+(2 N_\beta+1) \sinh_{2 r} \sin_\psi \right)\,,
	\end{aligned}
\end{equation}
where we denote the thermal expectation value (computed with \eqref{thermalstate}) of the number operator as
\begin{equation}\label{Nbeta}
	N_\beta :=	\langle \hat{N} \rangle_{thermal} = \tr \left( \hat{\rho}_\beta \, \hat{a}^\dagger\hat{a} \right)= \frac{1}{e^{\beta } -1} \,.
\end{equation}
By means of \eqref{Nbeta}, the reduction to pure states ($\beta\rightarrow\infty$) is achieved by setting $N_\beta=0$. Next, we evaluate variances \eqref{variance} and covariances \eqref{covariance}. Incorporating the displacement and squeezing phases into a shorthand $\mathcal{F}_\pm =\cos_{2 \vartheta } \cos_\psi \pm \sin_{2 \vartheta } \sin_\psi$ and noticing that $2N_\beta +1= \coth_{\beta/2}$, one finds
\begin{widetext}
\begin{equation}\label{Var}
	\begin{aligned}
		(\Delta \hat{V})_\text{G}^2 &=	 \frac{v^2}{4}  \left(4 | \alpha | ^2 \coth_{\frac{\beta}{2}} \left( \cosh_{2 r}+ \mathcal{F}_+ \sinh_{2 r}\right)+\coth_{\frac{\beta}{2}}^2 \cosh_{4 r}-1\right) \,,  \\
		(\Delta \hat{H})_\text{G}^2  &= \frac{\omega ^2}{8}  \left( 8|\alpha|^2 \coth_{\frac{\beta}{2}} \left( \cosh_{2 r}+  \mathcal{F}_-\sinh_{2 r}\right)+\coth_{\frac{\beta}{2}}^2 \left(1+ 2 \sinh ^2_{2 r} \cos_{2 \psi} +\cosh_{4 r}\right)+2\right) \,, \\
		(\Delta \hat{C})_\text{G}^2  &= \frac{v^2}{8}  \left(8 | \alpha | ^2 \coth_{\frac{\beta}{2}} \left( \cosh_{2 r}-\mathcal{F}_- \sinh_{2 r}\right)+\coth_{\frac{\beta}{2}}^2 \left(1+2 \cosh_{4 r} \sin ^2_\psi +\cos_{2 \psi}\right) +2\right) \,,
			\end{aligned}
\end{equation}
and
\begin{equation}\label{Cov}
	\begin{aligned}
		\Delta( \hat{V}\hat{H})_\text{G} &= -\frac{v \omega}{4} \left( 	 4 |\alpha |^2\coth_{\frac{\beta}{2}}\left(\sinh_{2 r}\cos_\psi + \cosh_{2 r} \cos_{2 \vartheta}\right)+\coth_{\frac{\beta}{2}}^2 \sinh_{4 r} \cos_\psi \right) \,,\\
		\Delta(  \hat{V}\hat{C})_\text{G}  &=  \frac{v^2}{4}  \left(4 |\alpha|^2 \coth_{\frac{\beta}{2}} \left(\sinh_{2 r} \sin_\psi + \cosh_{2 r}\sin_{2\vartheta} \right)+\coth_{\frac{\beta}{2}}^2 \sinh_{4 r} \sin_{\psi} \right) \,,\\
		\Delta( \hat{H}\hat{C})_\text{G} & =  - \frac{v \omega}{4}   \left(4     |\alpha|^2 \coth_{\frac{\beta}{2}} \sinh_{2 r} (\sin_{2\vartheta}\cos_\psi + \cos_{2 \vartheta}\sin_\psi) +  \coth_{\frac{\beta}{2}}^2 \sinh^2_{2 r} \sin_{2 \psi} \right)  \,.
	\end{aligned}
\end{equation}
\end{widetext}
The quantities in \eqref{EV}, \eqref{Var} and \eqref{Cov} combine in a nontrivial way coherent, squeezed and thermal contributions; they generalise the expressions for simple states reported in appendix \ref{UncertApp}, being now (at the same time) functions of $\alpha = |\alpha|e^{{\rm i} \vartheta}$, $z=r e^{{\rm i}\psi}$ and $\beta$.

Expectation values, variances and covariances are all the ingredients one needs to analyse the semiclassical criteria discussed in section \ref{StatesSection}. For instance, using the first two expressions in \eqref{Var}, the first in \eqref{Cov} and the last in \eqref{EV}, it is straightforward (albeit tedious) to see that the Robertson--Schr\"odinger uncertainty principle \eqref{RSprinciple} is \textit{not} minimised by Gaussian states at $\chi=0$. One can in fact prove that the inequality is never saturated for any $\chi$, much like with coherent states. As expected, the inequality becomes an identity only when $\alpha = N_\beta =0$, which is the case of a pure squeezed state \eqref{squeezeUR}. Details on the Robertson--Schr\"odinger principle for Gaussian states are given at the end of appendix \ref{UncertApp}.

More importantly, we now show that Gaussian states can be chosen to have small quantum fluctuations. Even with such a large parameter space (spanned by $\alpha$, $z$ and $\beta$), one can notice from \eqref{Var} and \eqref{EV} that it is always possible to manipulate the displacement parameter $\alpha$ to make relative uncertainties arbitrarily small at $\chi=0$. While squeezed and thermal states alone do not allow for such a feature, squeezing and thermal effects can lead to semiclassical Gaussian states as long as one uses a large enough displacement. To make this more explicit, we can expand the fluctuations stemming out of \eqref{Var} and \eqref{EV} for large $|\alpha|$, obtaining
\begin{align}
	\frac{(\Delta \hat{V})_\text{G}^2}{\langle \hat{V} \rangle_\text{G}^2}  &\;\sim \; \frac{1}{|\alpha|^2} \coth_{\frac{\beta}{2}} (\cosh_{2r} + \mathcal{F}_+\sinh_{2r}) \,, \label{VolRUG}\\
	\frac{(\Delta \hat{H})_\text{G}^2}{\langle \hat{H} \rangle_\text{G}^2} & \; \sim \;  \frac{1}{|\alpha|^2 \cos^2_{2\vartheta} } {\coth_{\frac{\beta}{2}} (\cosh_{2r} +\mathcal{F}_- \sinh_{2r})} \,, \label{HamRUG} \\
	\frac{(\Delta \hat{C})_\text{G}^2}{\langle \hat{C} \rangle_\text{G}^2}	&\;\sim \; \frac{1}{|\alpha|^2 \sin^2_{2\vartheta} } {\coth_{\frac{\beta}{2}} (\cosh_{2r} -\mathcal{F}_- \sinh_{2r})} \,. \label{CRUG}
\end{align}
These expressions still refer to $\chi=0$, so they generalise \eqref{RUCS0} and \eqref{RUSS0}. We now discuss the dynamics of quantum fluctuations, focussing on the volume operator.
\begin{figure*}[t]
	\begin{center}
		\includegraphics[width=.32\textwidth]{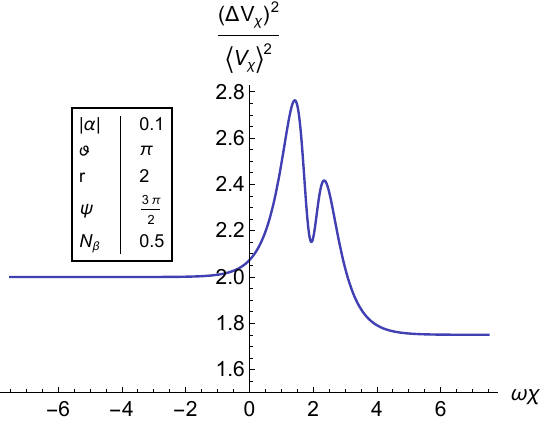}
		\hspace{0.1cm}
		\includegraphics[width=.32\textwidth]{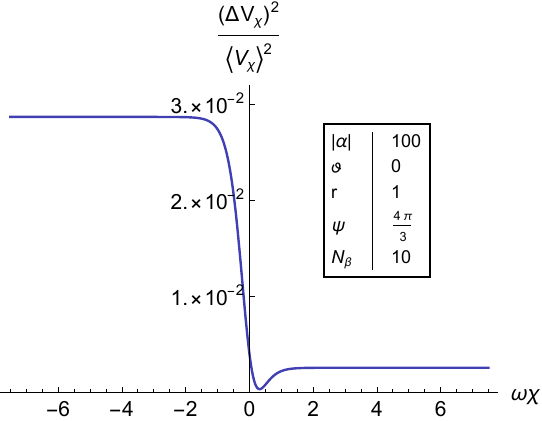}
		\hspace{0.1cm}
		\includegraphics[width=.32\textwidth]{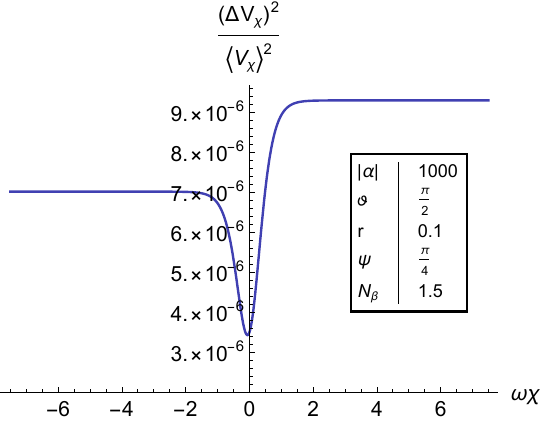}	
	\end{center}
	\caption{\small Volume relative uncertainties with Gaussian states for some state parameters.}
	\label{gsplots}
\end{figure*}

Recall from section \ref{gftcosmosection} that the single-mode GFT Hamiltonian \eqref{HV} makes the evolution operator $\hat{U}(\chi) = e^{-{\rm i} \hat{H} \chi}$ a squeezing operator (with purely imaginary squeezing parameter). Relations allowing a reordering of displacement and squeezing operators, or the composition of two squeezing operators into one, are well known (see appendix \ref{GaussThermoApp}), and it might be tempting to work in the Schr\"odinger picture and to define
\begin{equation}\label{Schro}
	\hat{\rho}_\text{G} (\alpha, z, \beta; \chi) = \hat{U}(\chi) \hat{D}(\alpha) \hat{S}(z) \hat{\rho}_\beta \hat{S}^\dagger(z) \hat{D}^\dagger(\alpha) \hat{U}^\dagger(\chi) \,,
\end{equation}
as a time-dependent Gaussian state. However, one finds that using properties such as \eqref{comp} and \eqref{closerules1} on \eqref{Schro} leads to very complicated calculations, due to the mixing of  parameters. We thus keep working in the Heisenberg picture, in which the explicit dynamical equations \eqref{vofchi}, \eqref{chivar} and \eqref{chicovar} allow us to obtain the $\chi$-evolution of all the quantities of interest. For example, the dynamics of the volume expectation value reads
\begin{equation}
	\begin{aligned}
	\langle \hat{V}_\chi \rangle_\text{G} &= 
	\frac{v}{2}\big[ 2|\alpha|^2 (\cosh_{2 \omega \chi  } + \sinh_{2 \omega \chi }\sin_{2\vartheta} ) -1   \\& + \coth_{\frac{\beta}{2}} ( \sinh_{2   \omega \chi}\sinh_{2 r} \sin_\psi   +\cosh_{2   \omega \chi}\cosh_{2 r} ) \big]\,	.
	\end{aligned}
\end{equation}
While it is not useful to show all the other $\chi$-dependent counterparts of \eqref{EV}, \eqref{Var} and \eqref{Cov} in full, one can compute them in the same fashion (also using \eqref{cofchi} for quantities containing $\hat{C}$). Crucially, we find that Gaussian states can be chosen to make the volume relative uncertainties, ${(\Delta \hat{V}_\chi)_{\text{G}}^2}/{\langle \hat{V}_\chi \rangle^2_{\text{G}}}$, arbitrarily small at all times. We can analytically show this again by considering the asymptotic behaviour of the evolving volume fluctuations for large $|\alpha|$, finding
\begin{widetext}
\begin{equation}\label{mainresult}
	\frac{(\Delta \hat{V}_\chi)_{\text{G}}^2}{\langle \hat{V}_\chi \rangle^2_{\text{G}}}
	\sim \frac{\coth_{{\beta}/{2}}}{|\alpha|^2}\frac{\sinh _{4   \omega\chi}  (\sin_{2 \vartheta }+\sinh_{2 r} \sin_\psi)+\cosh _{4   \omega\chi} (\sinh_{2 r}\sin_{2 \vartheta}   \sin_\psi +1)+ \sinh_{2 r} \cos_{2 \vartheta}   \cos_\psi }{(\sin_{2\vartheta} \sinh_{2   \omega\chi}+\cosh_{2   \omega\chi})^2}\,.
\end{equation}
\end{widetext}
Again avoiding the problematic values $\vartheta=\frac{\pi}{4}+k\frac{\pi}{2}$, this is the product of a bounded function (in $\chi$) and a factor $1/|\alpha|^2$, which can hence be made arbitrarily small at all times. The function exhibits similar features to the coherent-states case and generalises \eqref{largealphaCS} to Gaussian states by also keeping squeezing and thermal contributions. The dynamics hence do not spoil the semiclassical behaviour of suitably chosen Gaussian states. To give some graphical intuition, we also plot in figure \ref{gsplots} a few instances illustrating some interplay between the various state parameters, in particular exemplifying that $|\alpha|$ can make ${(\Delta \hat{V}_\chi)_\text{G}^2}/{\langle \hat{V} _\chi \rangle_\text{G}^2}$ as small as desired at all times.  Compared to the choice of $|\alpha|$, the other parameters seem to have relatively minor impact on the relative uncertainty.

Finally, we can explicitly find the asymptotic behaviour represented by the plateaus in figure \ref{gsplots},
\begin{widetext}
\begin{equation}\label{AGS}
	\frac{(\Delta \hat{V}(\chi))_{\text{G}}^2}{\langle \hat{V}(\chi) \rangle^2_{\text{G}}} \quad \overset{\chi\rightarrow \pm \infty}{\longrightarrow} \quad
2- \frac{8 | \alpha | ^4 (\sin_{2 \vartheta} \pm1)^2}{  \left[ 2| \alpha |^2 (\sin_{2 \vartheta} \pm 1) \pm \coth_{{\beta}/{2}} (\cosh_{2 r}\pm\sinh_{2 r} \sin_\psi )\right]^2} =:  \mathcal{A}_\text{G} \,,
\end{equation}
\end{widetext}
which generalises \eqref{ACS} and shares the same properties. In particular, approximating \eqref{AGS} for large $|\alpha|$ we can see that it can be made arbitrarily small:
\begin{equation}
	\mathcal{A}_\text{G}	\quad \sim \quad    \frac{2}{|\alpha|^2} \frac{ \coth_{{\beta}/{2}} (\cosh_{2 r}\pm\sinh_{2 r} \sin_\psi )}{ (1\pm\sin_{2 \vartheta} )} \,.
\end{equation}
We can also expand ${(\Delta \hat{V}_\chi)_\text{G}^2}/{\langle \hat{V} _\chi \rangle_\text{G}^2}$ in inverse volume powers as per \eqref{expansion}, finding
\begin{equation}
	\frac{(\Delta \hat{V}_\chi)^2_{\text{G}}}{\langle \hat{V}_\chi \rangle_{\text{G}}^2} =  \mathcal{A}_{\text{G}} \left(1+ \frac{v}{\langle \hat{V}_\chi \rangle}\right) +  \mathcal{C}_{\text{G}}\frac{v^2}{\langle \hat{V}_\chi \rangle^2} +  \mathcal{O}\left(\frac{1}{\langle \hat{V}_\chi \rangle^3}\right) \,,
\end{equation}
where, using $h_\pm= \coth_{\frac{\beta}{2}}(\cosh_{2 r}\pm \sinh_{2 r} \sin_\psi) $ to encapsulate squeezing and thermal contributions,
\begin{widetext}
\begin{equation}
	\begin{aligned}
		\mathcal{C}_{\text{G}} &= \frac{ h_\pm \left(4 |\alpha|^2 (\sin_{2 \vartheta}\pm1)\pm h_\pm\right)}{8} \Bigg[\frac{8|\alpha|^2 \cos_{2\vartheta} \cot_\psi (h_\pm-h_\mp)\pm \cot_\psi^2(h_\pm-h_\mp)^2	\mp4}{ 2 h_\pm \left(h_\pm \pm 4 |\alpha|^2 (\sin_{2 \vartheta}\pm1)\right)} \\& +\frac{   2 |\alpha|^2 (\sin_{2 \vartheta}\mp 1) \mp h_\mp }{h_\pm \pm 2 |\alpha|^2 (\sin_{2 \vartheta}\pm1)} \mp \frac{ 2|\alpha|^2\left( h_++h_-+
			2|\alpha|^2 \cos_{2 \vartheta}^2  \pm \sin_{2\vartheta} (h_\mp - h_\pm)   \right)  +h_- h_+ -4 }{\left(h_\pm \pm 2 |\alpha|^2 (\sin_{2 \vartheta}\pm 1)\right)^2}\Bigg] \,.
	\end{aligned}
\end{equation}
\end{widetext}
Interestingly, here again $\mathcal{B}_{\text{G}}= \mathcal{A}_{\text{G}}$, as with all the other states. One can check that $\mathcal{C}_{\text{G}}$ reduces to \eqref{CCS} for $r=0$ and $\beta\rightarrow\infty$ (since $h_\pm\rightarrow1$) and vanishes for $\alpha=0$ and $\beta\rightarrow\infty$ (cf.\ \eqref{RUSS}).

To summarise, we have seen that the general family of Gaussian states contains states with small relative uncertainties and can thus be regarded as semiclassical. {This} property is realised for the volume, the Hamiltonian and the $\hat{C}$ operator in the context of GFT cosmology, and it holds at all times (crucially at late times, where quantum fluctuations are actually expected to be small). Because all harmonic cosmologies \cite{Bojo_2019,HarC1,*HarC2} rely on the same underlying Lie algebra, these results actually hold for any general $\mathfrak{su(1,1)}$ (or CVH) {quantum} cosmological scenario. As for coherent states, Gaussian states do not saturate the Robertson--Schr\"odinger uncertainty principle, showing again that it is not clear whether such a criterion should be invoked to classify states as semiclassical. 

\section{Algebraic approach to GFT cosmology}\label{algebraicsection}

In this section we explore the role and behaviour of semiclassical states in the original formalism of the GFT cosmology programme \cite{Oriti_GFT2ndLQG}, which we denote algebraic approach, following \cite{Aniso}. In particular, we investigate whether generalised Gaussian states can exist, and we highlight differences with the deparametrised approach adopted in previous sections.

The algebraic formalism for group field theory shares some similarities with a standard Dirac quantisation. It is based on the construction of a kinematical Hilbert space of abstract states, among which \textit{physical} states are selected by demanding that they satisfy a constraint coming from the underlying theory. Following section \ref{gftcosmosection}, states in the GFT kinematical Fock space would be unphysical quantum tetrahedra (or spin network-like states), on which dynamical equations are imposed \textit{a posteriori}, usually in a mean-field regime \cite{Gielen_2016,GFTcosmoLONGpaper,Oriti_2016,*BOriti_2017}. However, unlike in a Dirac quantisation, one here assumes that physical states are elements of the original Hilbert space. This assumption is strictly speaking inconsistent since such states have infinite norm.\footnote{For a more standard Dirac quantisation of (free) GFT that uses group averaging to define a physical Hilbert space, see \cite{frozenf}. The resulting theory is close, though not exactly equivalent, to the deparametrised theory defined earlier.} Moreover, working in a ``timeless'' setting, one defines kinematical operators as relational observables (e.g., the volume as a function of the scalar field $\chi$), which may also contain infinities.  We will encounter divergences at many points in this section, so that our expressions need to be treated as formal and subject to some regularisation procedure (some ideas for dealing with these infinities include \cite{Isha_thermalGFT,*Isha_thermalGFT2,*IshaDaniele,Marchetti2021}). We are mainly interested in a general conceptual comparison with the analysis of previous sections; regardless of divergences one can check whether one can define Gaussian-like states in this formalism, in the sense of at least approximately physical states.

Aiming to extract effective cosmological dynamics, we work with a similar setup to the one described in section \ref{gftcosmosection}, the main difference being that the group field \eqref{groupfield} is now complex valued. Restricting again to a single Peter--Weyl mode (and hence dropping $SU(2)$ arguments as in section \ref{Subsu11FLRW}), we deal with a free complex group field theory, where the classical equation of motion reads
\begin{equation}\label{eom}
	\left(\partial^2_\chi -\omega^2 \right) \varphi(\chi) =0
\end{equation}
with $\omega^2 = -K^{(0)}/K^{(2)}$ (cf.\ \eqref{freeGFT}). A canonical quantisation performed in this approach requires promoting the field and its complex conjugate to operators $\hat{\varphi}$ and $\hat{\varphi}^\dagger$ with
\begin{equation}\label{algcom}
	\left[\hat{\varphi}(\chi),\hat{\varphi}^\dagger(\chi')\right] = \delta (\chi-\chi') \,.
\end{equation}
The key deviation from the deparametrised approach of section \ref{gftcosmosection} is the fact that the operators $\hat{\varphi}(\chi)$ and $\hat{\varphi}^\dagger(\chi)$ are not thought of as evolving in time, but as separate independent operators for each value of $\chi$. Using these as abstract ladder operators, one can now construct a kinematical Fock space with vacuum $|0\rangle$ satisfying $\hat{\varphi}(\chi)|0\rangle=0$. Then, for instance, the state associated with a single (unphysical) GFT quantum with a given (fixed) value of $\chi$ reads $\hat{\varphi}^\dagger(\chi) |0\rangle$, not to be confused with the dynamical one-particle state $\hat{a}^\dagger(\chi) |0\rangle$ of section \ref{Subsu11FLRW}. Kinematical analogues of the number and volume operators are naturally defined in this approach as $	\hat{V}(\chi)=   v \, \hat{N}(\chi)  =  v \, \hat{\varphi}^\dagger(\chi) \hat{\varphi} (\chi) $.

Among these kinematical states, one needs to identify physical ones. To do this, from the GFT action one can derive Schwinger--Dyson equations for correlation functions,
\begin{equation}\label{SDeq}
	\left< \Psi \left| \frac{\delta \hat{\mathcal{O}}} {\delta \hat{\varphi}^\dagger} - \hat{\mathcal{O}} \frac{\delta S[\hat{\varphi},\hat{\varphi}^\dagger] }{\delta \hat{\varphi}^\dagger} \right| \Psi  \right> =0 \,,
\end{equation}
which must be satisfied for physical states, where $\hat{\mathcal{O}}$ is any polynomial functional of the field operators. In practice, one truncates this infinite tower of equations by considering a few very simple choices for $\hat{\mathcal{O}}$, the most common being the identity operator. Indeed setting $\hat{\mathcal{O}}=1$ in \eqref{SDeq} amounts to requiring that the operator version of the Euler--Lagrange equations hold on average. 

Alternatively, one can require physical states to satisfy \cite{Gielen_2016,GFTcosmoLONGpaper,Oriti_2016,*BOriti_2017}
\begin{equation}\label{constraint}
	\frac{\delta S[\hat{\varphi},\hat{\varphi}^\dagger]}{\delta \hat{\varphi}^\dagger} \bigg|\, \Psi \bigg> = 0     \,,
\end{equation}
which can be seen as the imposition of a quantum constraint on $|\Psi\rangle$ typical of a Dirac quantisation scheme.  Compared to requiring some expectation value to vanish (usually given by the first Schwinger--Dyson equation), \eqref{constraint} provides a stronger condition for defining exactly physical states. We shall see that one can quickly find at least one exact solution for a free GFT; moreover, we explore some other possibilities in appendix \ref{AppCondensates}, where we show that to be a solution of the constraint \eqref{constraint} (and hence physical), a generic $|\Psi\rangle$ must satisfy strict conditions. Whether one uses \eqref{constraint} or a truncation of \eqref{SDeq}, the task in this approach is to find equations for the functions defining the states. Such conditions ensure the corresponding states are either exact or approximate solutions of the quantum dynamics.

The only class of states (including the more specific proposal of \cite{Marchetti2021}) which has been successfully used to extract cosmological dynamics is given by field coherent states
\begin{equation}\label{sigma}
	\begin{aligned}
	|\sigma \rangle  & = \hat{\mathcal{D}}(\sigma) |0\rangle\,, \\ \hat{\mathcal{D}}(\sigma) &= \exp\left( \int {\rm d} \chi  \left[\sigma(\chi)\hat{\varphi}^\dagger(\chi) - \overline{\sigma(\chi)} \hat{\varphi}(\chi)\right] \right)\,,
	\end{aligned}
\end{equation}
which we define in analogy with \eqref{CS} using a displacement-like operator. Due to the Baker--Campbell--Hausdorff formula, \eqref{sigma} is equivalent to a single-particle condensate state of the type usually adopted in the literature, $	|\sigma \rangle = \mathcal{N}_\sigma \exp\left(\int {\rm d} \chi \, \sigma (\chi) \hat{\varphi}^\dagger(\chi) \right)|0\rangle$ with $ \mathcal{N}_\sigma = \exp\left( -\frac{1}{2} \int {\rm d} \chi |\sigma(\chi)|^{2} \right)$, which explicitly shows a divergent norm ($\mathcal{N}_\sigma =0$ for any $\sigma$ solving \eqref{eom}). One way of regularising the state $|\sigma\rangle$ is by introducing an \textit{ad hoc} cutoff in $\chi$, which would represent an arbitrarily large (but finite) range of validity for the resulting effective relational dynamics.\footnote{One might argue that this does not represent an issue as the free-theory approximation breaks down at some $\chi$ (when the interactions in \eqref{GFTaction} become important), so one should not trust the model for too large $\chi$ anyway.}

While often described as an approximate solution in the literature (even for GFT models based on the free theory), we stress that states of the form \eqref{sigma} can solve \eqref{constraint} \textit{exactly}. Indeed, due to the property $ \hat{\varphi} (\chi) |  \sigma  \rangle =  \sigma   (\chi )|\sigma \rangle $, \eqref{sigma} is a physical state provided that the displacement parameter $ \sigma   (\chi )$ satisfies the classical free GFT equation of motion \eqref{eom}, namely $\left(\partial^2_\chi - \omega^2 \right) \sigma   (\chi )=0$. The solution to this equation dictates how dynamics are implemented in the algebraic approach, as geometrical quantities inherit $\chi$-dependence through $ \sigma   (\chi )$. In particular, one can obtain the volume expectation value
\begin{equation}\label{volumeALG}
	\begin{aligned}
	\langle \hat{V}(\chi) \rangle _\sigma &= v \frac{\langle \sigma | \hat{\varphi}^\dagger(\chi) \hat{\varphi} (\chi)  |\sigma\rangle}{\langle \sigma | \sigma \rangle } =  v |\sigma (\chi) |^2 \,, \\ \sigma (\chi) &= A e^{\omega \chi} + B e^{-\omega \chi}\,,
	\end{aligned}
\end{equation}
where $A$ and $B$ are constants, and show that it satisfies a Friedmann-like equation similar to \eqref{Friedmann}. This can be expressed by means of the quantities \cite{Oriti_2016,*BOriti_2017,Gielen_lowspin}
\begin{equation}\label{quantities}
%	\begin{aligned}
	E = -4 \text{Re} (A\overline{B})\,,  \qquad  Q = 2 \text{Im} (A\overline{B})\,,
%	\end{aligned}
\end{equation}
so that
\begin{equation}\label{FriedmannSigma}
	\left(\frac{1}{\langle \hat{V}(\chi)\rangle_\sigma } \frac{{\rm d} \langle \hat{V}(\chi)\rangle_\sigma }{{\rm d} \chi}\right)^2 = 4\omega^2 \left( 1 +\frac{ v E}{\langle \hat{V}(\chi)\rangle_\sigma } 	-\frac{ v^2 Q^2}{ \langle \hat{V}(\chi)\rangle_\sigma ^2} \right) 	  \,.
\end{equation} 
Apart from differences in numerical factors, such effective cosmological dynamics share the properties of \eqref{Friedmann} and essentially describe the same scenario discussed at the end of section \ref{Subsu11FLRW}. The crucial difference is that in order to obtain \eqref{FriedmannSigma} we had to specifically use the state \eqref{sigma}, as indicated by the index of $\langle \hat{V}(\chi) \rangle _\sigma$. In contrast, the effective Friedmann equation \eqref{Friedmann} of the deparametrised approach holds in any state.

Note that even if the volume itself does not show any infinities, volume fluctuations diverge as
\begin{equation}\label{deltasigma}
	\frac{(\Delta \hat{V})_\sigma^2}{\langle\hat{V} \rangle^2 _\sigma} =\frac{\delta(0)}{|\sigma(\chi)|^2}\,.
\end{equation}
If one removes the distribution $\delta(0)$ by means of some regularisation procedure (e.g., replacing the Dirac delta in \eqref{algcom} with a Kronecker delta by considering smeared observables \cite{Isha_thermalGFT,*Isha_thermalGFT2,*IshaDaniele}, or imposing the dynamics in a different way by working with the peaked coherent states of \cite{Marchetti2021}), \eqref{deltasigma} gets automatically smaller and smaller over time as $\sigma(\chi)$ grows exponentially (cf.\ \eqref{volumeALG}). In this sense, one might then argue that these coherent states are semiclassical.

Because the effective Friedmann equation \eqref{FriedmannSigma} seems to rely on coherent states, it is natural to ask whether one can use more general states, such as Gaussian states, to obtain a similar result. Given how dynamics are implemented in the algebraic approach, we shall see that it is not clear whether Gaussian states are a useful option for this framework. In order to define generalised Gaussian states we resort to the thermofield formalism since a well-defined procedure in terms of thermal-like density matrices is not directly available in the algebraic approach to GFT. The thermofield dynamics were developed in the context of GFT in \cite{Isha_thermalGFT,*Isha_thermalGFT2,*IshaDaniele} for thermal coherent states; this naturally extends to the case of Gaussian states following the strategy of appendix \ref{GaussThermoApp}, with suitably generalised definitions. Explicitly, in analogy with \eqref{thermalGauss}, a Gaussian-like state in the algebraic approach can be defined as\footnote{Just like \eqref{sigma}, the state \eqref{GSalgebraic} has a divergent norm. Again, one could impose a cutoff in the $\chi$ integrations, but in this case the divergences are even more severe. See appendix \ref{AppCondensates} for details on condensate states and their norms.}
\begin{equation}\label{GSalgebraic}
	|\sigma, \zeta, \beta \rangle =	\hat{\mathcal{D}}(\sigma) \hat{\mathcal{S}}(\zeta) |0_\beta\rangle\,,
\end{equation}
where we introduce a squeezing-like operator  
\begin{equation}\label{squeezingALG}
	\hat{\mathcal{S}}(\zeta) = \exp\left( \frac{1}{2}\int {\rm d} \chi \left[\zeta(\chi) \hat{\varphi}^{\dagger 2}(\chi) - \overline{\zeta(\chi) } \hat{\varphi}^2(\chi)\right] \right)\,,
\end{equation}
and the algebraic counterpart of the thermal vacuum \eqref{thvacuum},
\begin{equation}
	\begin{aligned}
	|0_\beta \rangle &= \hat{\mathcal{T}} (\theta_\beta) |0,\tilde{0}\rangle \,, \\ \hat{\mathcal{T}} (\theta_\beta) &= \exp\left( \int {\rm d} \chi \, \theta_\beta(\chi) \left[ \hat{\varphi}^{\dagger}(\chi) \hat{\tilde{\varphi}}^{\dagger}(\chi)  -  \hat{\varphi}(\chi) \hat{\tilde{\varphi}}(\chi)\right] \right)\,.
	\end{aligned}
\end{equation}
The general construction of appendix \ref{GaussThermoApp} applies here: the state $|0,\tilde{0}\rangle$ is a product vacuum for the doublet \textit{kinematical} Hilbert space, the tilde operators $\hat{\tilde{\varphi}}(\chi)$ and $\hat{\tilde{\varphi}}^{\dagger}(\chi)$ satisfy the algebra \eqref{algcom} while commuting with non-tilde operators, \textit{et cetera}. In particular, one can make use of the relation $\sinh^2(\theta_\beta(\chi)) = (e^{\beta(\chi)}-1)^{-1}$ (cf.\ \eqref{thermomap}) to express results in terms of the statistical parameter $\beta(\chi)$.

Such a Gaussian-like state is not physical as it does not solve the constraint \eqref{constraint}. More precisely, focussing on the case of a pure Gaussian state for simplicity, one can start from \eqref{constraint} with $|\sigma, \zeta \rangle =	\hat{\mathcal{D}}(\sigma) \hat{\mathcal{S}}(\zeta) |0\rangle$ and obtain the condition 
{\small \begin{equation}\label{gsunph}
	\left(\partial^2_\chi - \omega^2 \right) \left(\sigma(\chi) + \frac{\zeta (\chi)}{|\zeta (\chi)|} \tanh (|\zeta (\chi)|)  \left(\hat{\varphi}^\dagger(\chi) - \overline{\sigma(\chi)}\right)\right) \\ =0\,,
\end{equation}}
which cannot generically be solved for the displacement and squeezing functions $\sigma(\chi)$ and $\zeta(\chi)$. Including thermal contributions only results in a more complicated equation with no interesting novelties, as $\hat{\mathcal{T}}(\theta_\beta)$ is essentially a generalised squeezing operator just like $\hat{\mathcal{S}}(\zeta)$. While setting $\zeta=0$ in \eqref{gsunph} returns the classical equation of motion for $\sigma(\chi)$ (which makes the coherent-like state \eqref{sigma} physical), notice that setting $\sigma=0$ shows that a purely squeezed-like state $|\zeta\rangle=\hat{\mathcal{S}}(\zeta) |0\rangle$ is also unphysical as
\begin{equation}\label{squnotphy}
	\left(\partial^2_\chi - \omega^2 \right) \left(\frac{\zeta (\chi)}{|\zeta (\chi)|} \tanh (|\zeta (\chi)|) \,\hat{\varphi}^\dagger(\chi)\right) =0
\end{equation} 
cannot yield a solution for the squeezing function $\zeta(\chi)$. 

The usual strategy in this situation is to shift the attention towards averages, and require that the state be only approximately physical. We can use the general Gaussian states \eqref{GSalgebraic} in \eqref{SDeq}, e.g.\ with $\hat{\mathcal{O}}=1$, to determine the form of our state parameters as functions of $\chi$. One finds that the first Schwinger--Dyson equation does not provide a condition for the squeezing and thermal functions:
\begin{equation}\label{averageGaussian}
	\begin{aligned}
	\left< \frac{\delta S[\hat{\varphi},\hat{\varphi}^\dagger]}{\delta \hat{\varphi}^\dagger} \right>_{\sigma,\,\zeta,\,\beta} & = \left(\partial^2_\chi -\omega^2 \right) \langle\hat{\varphi}(\chi) \rangle_{\sigma,\,\zeta,\,\beta} \\& =
\left(\partial^2_\chi -\omega^2 \right) \sigma (\chi) =0 \,.
	\end{aligned}
\end{equation}
As essentially observed in \cite{Isha_thermalGFT,*Isha_thermalGFT2,*IshaDaniele} for thermal coherent states, we can only determine the $\chi$-dependence for the displacement parameter $\sigma(\chi)$, which in particular is the same as in the (pure) coherent-states scenario. This is due to the fact that squeezed states $|\zeta\rangle$, and thus also squeezed thermal states $|\zeta,\beta \rangle$, have a vanishing field expectation value $\langle \zeta, \beta | \hat{\varphi}(\chi) |\zeta,\beta \rangle =0$. As a consequence, one cannot use \eqref{averageGaussian} to extract dynamical information for $\zeta$ and $\beta$.

Since going to Schwinger--Dyson equations of higher order is rather complicated (see appendix \ref{AppCondensates} for an attempt with dipole states and squeezed-like states), we can follow the idea of \cite{Isha_thermalGFT,*Isha_thermalGFT2,*IshaDaniele} and assume that the parameters $\zeta$ and $\beta$ are constant. While the dynamics is still governed by the same function $\sigma(\chi)$, this simple generalisation does affect observable averages (such as the volume), and hence the resulting Friedmann equation, with new \textit{static} contributions of squeezing and thermal nature. From the volume expectation value computed with \eqref{GSalgebraic},
\begin{widetext}
\begin{equation}
%	\begin{aligned}
	\langle \hat{V}(\chi)  \rangle _{\sigma,\,\zeta,\,\beta} = v  \Big[|\sigma(\chi)|^2 + \delta(0)\sinh^2(|\zeta|)  +   \delta(0)  (e^\beta -1)^{-1}\cosh(2|\zeta|) \Big] \,, 
%	\end{aligned}
\end{equation}
one finds the following effective Friedmann equation:
\begin{equation}\label{GaussFE}
		\left(\frac{1}{\langle \hat{V}(\chi)\rangle _{\sigma,\,\zeta,\,\beta} } \frac{{\rm d} \langle \hat{V}(\chi)\rangle _{\sigma,\,\zeta,\,\beta} }{{\rm d} \chi}\right)^2 = 4\omega^2\left(  1 +\frac{ v }{\langle \hat{V}(\chi)\rangle _{\sigma,\,\zeta,\,\beta} } (E+\mathcal{E}) 	-\frac{ v^2 }{ \langle \hat{V}(\chi)\rangle _{\sigma,\,\zeta,\,\beta} ^2} (Q^2+\mathcal{Q}^2)  \right) \,,
	\end{equation}
\end{widetext}
where $E$ and $Q$ are given in \eqref{quantities} and the squeezing and thermal contributions are encoded in
\begin{equation}\label{Newquantities}
	\begin{aligned}
	\mathcal{E} &= \delta(0)\big[1- \coth(\beta/2)\cosh (2 |\zeta|) \big]\,, \\ \mathcal{Q}^2 &= - \frac{1}{4}\mathcal{E} \left(\mathcal{E}+2 E \right)\,.
	\end{aligned}
\end{equation}
As mentioned, we formally keep the Dirac delta distributions in these expressions assuming one can get rid of them by using, e.g., a smearing procedure \cite{Isha_thermalGFT,*Isha_thermalGFT2,*IshaDaniele}. Notice that such divergences affect the thermal and squeezing contributions already at the level of the volume expectation value.

Of course, when $\zeta=0$ and $\beta \rightarrow\infty$, \eqref{GaussFE} reduces to \eqref{FriedmannSigma} since $\mathcal{E} = \mathcal{Q} =0$. Similarly, one can check that the result of \cite{Isha_thermalGFT,*Isha_thermalGFT2,*IshaDaniele} with a ``static thermal cloud'' emerges by setting $\zeta=0$. We remark however that both \eqref{GaussFE} and the modified Friedmann-like equation of \cite{Isha_thermalGFT,*Isha_thermalGFT2,*IshaDaniele} represent only a somewhat weak generalisation of \eqref{FriedmannSigma} as the new contributions are assumed to be $\chi$-independent; this is an arbitrary assumption that was made because the model is not predictive for $\zeta$ and $\beta$ (cf.\ \eqref{averageGaussian}).

Along the same lines, one can also find new constant contributions to the volume fluctuations for general Gaussian-like states. Since all the $\chi$-dependence is encoded in the displacement parameter $\sigma(\chi)$, one can proceed to remove the divergences of the usual kind and find again that relative uncertainties are automatically tamed under time evolution. To give a concrete but concise example, pure\footnote{One can write down an expression including $\beta$, but it is not insightful to show this in full since thermal contributions are of squeezing type (see thermofield formalism in appendix \ref{GaussThermoApp}) and assumed to be constant.} Gaussian-like states $|\sigma,\zeta\rangle$ yield the following relative uncertainties
\begin{widetext}
	\begin{equation}\label{deltagaussbis}
	\frac{(\Delta \hat{V})_{\sigma,\,\zeta}^2}{\langle\hat{V} \rangle^2 _{\sigma,\,\zeta}} = \delta(0) \frac{2 |\zeta| |\sigma(\chi)|^2 \cosh (2 |\zeta|)+\sinh (2 |\zeta|) \left(\sigma (\chi)^2 \, \overline{\zeta}+ \overline{\sigma(\chi)}^2 \, \zeta\right) +\delta(0) |\zeta| \sinh^2 (2 |\zeta|)}{2 |\zeta| \left(|\sigma(\chi)|^2+ \delta(0) \sinh ^2(|\zeta|)\right)^2}\,.
\end{equation}
\end{widetext}
Since $\sigma(\chi)$ grows exponentially, one easily finds that at late times \eqref{deltagaussbis} reduces to an analogue of \eqref{VolRUG} (in this case with $\beta\rightarrow\infty$). Of course, in the limit $\zeta\rightarrow 0$ one recovers \eqref{deltasigma}. Notice that when $\sigma=0$ one is left with an expression which incidentally has no divergences, namely $	{(\Delta \hat{V})_{\zeta}^2}/{\langle\hat{V} \rangle^2 _{\zeta}}=2\coth^2 (|\zeta|)$ (as in \eqref{RUSS0}). We show a similar feature for dipole states (expected to be a type of squeezed state) in appendix \ref{AppCondensates}.

\section{Conclusions}

In this paper we constructed the wide class of (mixed) Gaussian states in the context of group field theory {and} $\mathfrak{su(1,1)}$ quantum cosmological models, and we analysed relevant properties for such states to be semiclassical and lead to a macroscopic cosmology. Inspired by bosonic theories in other areas of quantum mechanics, in particular quantum optics and quantum information theory, we defined the family of Gaussian states in its most general form, obtaining the most general state by applying displacement and squeezing operators to a pure thermal state. Since such states are generally defined as equilibrium (or Gibbs) states, Gaussian states should carry a statistical interpretation. In a discrete setting such as GFT, the notion of particle number allows one to think of a statistical ensemble where adding or removing quanta might come with some intrinsic cost in ``energy'' (usually the chemical potential in thermodynamics). In this sense, the ``thermal'' effects appearing in our results are not inherently related to some physical temperature like the one adopted in standard cosmology, but a grandcanonical interpretation might be more meaningful. From this perspective, discrete GFT models can add new features to cosmological scenarios which are missing in the continuum. However, giving a precise physical meaning to statistical parameters is highly nontrivial for a background-independent quantum gravity theory, where spacetime is only seen as emergent. Here we restrict ourselves to studying thermality from a mathematical point of view.  

While all existing work in GFT cosmology focuses on some type of coherent states, our results extend this past work to the most general family of semiclassical states, under the assumptions of a free theory and a single field mode. The general family includes pure coherent and squeezed states as special cases, as well as thermal (or generally mixed) states constructed along the lines of \cite{Isha_thermalGFT,*Isha_thermalGFT2,*IshaDaniele}. With the exception of section \ref{algebraicsection}, the paper mainly focuses on the deparametrised approach to canonical quantisation of GFT. We gave explicit analytical results for the relevant quantum fluctuations at all relational times, for both Fock coherent states and the Perelomov--Gilmore family of $SU(1,1)$ coherent states (herein called squeezed states). These return and generalise the late-time results of \cite{Gielen_2020}. We distinguish between two possible properties that can be associated with semiclassicality: the requirement that relative uncertainties (in the geometrical observables of interest) be small, and the saturation of the Robertson--Schr\"odinger uncertainty principle, which was never properly investigated in GFT (for similar work in loop quantum cosmology see, e.g., \cite{fraser}). We argue that saturation of the Robertson--Schr\"odinger equality is a less useful criterion in practice; for instance, simple Fock coherent states would not be semiclassical according to this, yet they yield relative uncertainties that can be arbitrarily small, which seems physically more relevant. On the other hand, states which do saturate the Robertson--Schr\"odinger inequality, such as squeezed states, show large relative uncertainties and are hence not semiclassical (as already argued in \cite{Gielen_2020}).

For a free GFT and single field mode, the dynamics are given by a simple quadratic bosonic Hamiltonian. Hence, the class of Gaussian states studied here represents the most general family of semiclassical states for the cosmological models under investigation. In the most general case, we studied the two possible semiclassicality criteria using analytical expressions for all the dynamical quantities of interest. One key result is that, while general Gaussian states just like coherent states do not saturate the Robertson--Schr\"odinger inequality, the fluctuations of the relevant geometrical observables (energy and volume) can be made arbitrarily small, at all relational times (especially in the late-time limit, where a classical cosmology is expected to emerge). Not too surprisingly, this behaviour is mainly governed by the displacement parameter, which can be manipulated arbitrarily so as to allow for nonvanishing thermal and squeezing contributions. In this sense, Gaussian states can be regarded as semiclassical. 

{All} such results rely on the algebra generated by the volume, the Hamiltonian and the $\hat{C}$ operator (dubbed CVH algebra in \cite{CVHalgebra}). For this reason, our findings apply to any general quantum system based on the $\mathfrak{su(1,1)}$ algebra, in particular provided that one can give a cosmological interpretation to some of the generators. Apart from GFT \cite{Gielen_2020}, a standard example of $\mathfrak{su(1,1)}$ cosmology is provided by loop quantum cosmology \cite{su11LQC,Boden1,*Boden2,*Boden3}, another realisation of the harmonic cosmology of \cite{Bojo_2019, HarC1,*HarC2}. Similar ideas were also applied to FLRW cosmological models in \cite{CFT1}, where the CVH algebra $\mathfrak{sl}(\mathfrak{2},\mathbb{R}) (\simeq \mathfrak{su(1,1)})$ allows one to formulate quantum cosmology as a one-dimensional conformal field theory.

We then studied similar questions in the algebraic approach to GFT cosmology, where states in a kinematical Hilbert space are subject to additional dynamical equations (constraints) in order to be seen as physical. In our attempts to generalise previous results to a wider class of states, we looked at different possible definitions of Gaussian-like states and used these to extract effective dynamics. A state is physical in an exact sense when it is annihilated by the free GFT equation of motion, and it can be considered approximately physical when at least one of the Schwinger--Dyson equations is satisfied. Ignoring as much as possible the known issues related to divergent norms, which affect all physical states in this formalism, we defined a family of generalised Gaussian states assuming that they could be regularised. Regardless of these divergences, we found that in the general family of Gaussian states there seem to be no (even approximately) physical states. The only Gaussian states that are physical turn out to be the pure coherent states considered in previous literature. In light of these findings, we analysed a simplified scenario, which includes the thermal construction of \cite{Isha_thermalGFT,*Isha_thermalGFT2,*IshaDaniele}, where all the time dependence is encoded in the displacement function. With this assumption we found an effective Friedmann equation which generalises previous works with new (albeit constant) contributions, as well as volume fluctuations that decrease over time, just like for coherent states. We also found a class of physical states that are more general than coherent states, but these are not Gaussian states and are (in general) not semiclassical.

In line with the ideas of exploring the connection between entanglement, entropy and geometry, which has seen recent attention for example in LQG \cite{Bianchi_Architecture,*Etera_Entanglement,*GluingLQG}, a natural direction for future work is to investigate similar questions in the context of GFT. In particular, it would be interesting to generalise the construction of our paper to multiple modes, specifically to more general squeezed states like those used in areas such as quantum optics \cite{Walls_Squeezing} or cosmology \cite{Grishchuk_SqueezeCosmo}. One could explore whether our results hold for two-mode (or generically multi-mode) Gaussian states, at least in the deparametrised approach where such states can be easily defined. This extension could add new features to GFT cosmological scenarios which are not captured by our single-mode construction.

\begin{acknowledgments}
This work was supported by the Royal Society through the University Research Fellowship Renewal URF$\backslash$R$\backslash$221005 (SG). 
\end{acknowledgments}

\appendix

%\begin{widetext}

\onecolumngrid

\section{Uncertainty principle}\label{UncertApp}

In this appendix we analyse in detail the Robertson--Schr\"odinger (RS) inequality \eqref{RSprinciple} for all states described in this paper. First, table \ref{TabCohSqu} reports the expectation values, variances and covariances for the operators of interest at $\chi=0$, using coherent states \eqref{CS} and squeezed states \eqref{SS}.
\begin{table}[b]
	\renewcommand{\arraystretch}{1.2}
	\begin{tabular}{p{3cm}p{6.5cm}p{5cm}}
		\toprule
		& Coherent states  & Squeezed states  \\
		\midrule
		$	\langle \hat{V}\rangle $ & $ v |\alpha|^2 $ & $v \sinh^2_r$\\
		\addlinespace
		$	\langle \hat{H}\rangle $ & $ - \frac{\omega}{2}\left(\bar{\alpha}^2 + \alpha^2 \right) = - \omega |\alpha|^2 \cos_{2\vartheta} $ & $- \frac{\omega}{2} \sinh_{2r} \cos_\psi$\\
		\addlinespace
		$	\langle \hat{C}\rangle $ & $ {\rm i}\frac{v}{2}\left(\bar{\alpha}^2 - \alpha^2 \right) = v |\alpha|^2 \sin_{2\vartheta} $ & $\frac{v}{2} \sinh_{2r} \sin_\psi $\\
		\addlinespace
		$(\Delta \hat{V})^2$ & $v^2 |\alpha|^2  $& $\frac{v^2}{2}  \sinh^2_{2r}$ \\
		\addlinespace
		$(\Delta \hat{H})^2$ & $\frac{\omega^2}{2}\left(1+ 2 |\alpha|^2 \right) $ & $\frac{\omega^2}{8} \left(3+\cosh_{4r} + 2 \cos_{2\psi} \sinh^2_{2r}\right)$ \\
		\addlinespace
		$(\Delta \hat{C})^2$ & $\frac{v^2}{2}\left(1+ 2 |\alpha|^2 \right)$& $\frac{v^2}{8} \left(3+\cosh_{4r} - 2 \cos_{2\psi} \sinh^2_{2r}\right)$\\
		\addlinespace
		$\Delta (\hat{V}\hat{H})$ & $- \frac{v \omega}{2} \left(\bar{\alpha}^2 + \alpha^2 \right) = - v \omega |\alpha|^2 \cos_{2\vartheta}  $& $- \frac{v \omega}{4}\cos_\psi \sinh_{4r} $ \\
		\addlinespace
		$\Delta (\hat{V}\hat{C})$ & ${\rm i}\frac{v^2}{2} \left(\bar{\alpha}^2- \alpha^2 \right) = v^2 |\alpha|^2 \sin_{2\vartheta}$&  $ \frac{v^2}{4} \sin_\psi \sinh_{4r}  $  \\
		\addlinespace
		$\Delta (\hat{H}\hat{C})$ & $0$& $- \frac{v \omega}{4} \sin_{2\psi} \sinh^2_{2r}$ \\
		\bottomrule
	\end{tabular}\caption{ \small Useful quantities computed with coherent states $|\alpha\rangle$ and squeezed states $|z\rangle$, where the displacement and the squeezing parameters are decomposed as $\alpha = |\alpha| e^{{\rm i} \vartheta}$ and $z = r e^{{\rm i} \psi}$. }	\label{TabCohSqu}
\end{table}
Moreover, using the analytical expressions for the time evolution of variances and covariances (cf.\ \eqref{chivar} and \eqref{chicovar}) we explicitly compute the dynamical behaviour of the left-hand side and right-hand side of the RS inequality \eqref{RSprinciple}. For coherent states we find
%\begin{widetext}
\begin{align*}\label{RSCSchi}
	{(\Delta \hat{V}_\chi )_{\text{C}}^2}{(\Delta \hat{H})_{\text{C}} ^2} &= \frac{v^2 \omega^2}{8} \left(2 |\alpha|^2+1\right) \left(4 |\alpha|^2 (  \sinh_{4   \omega \chi} \sin_{2 \vartheta}+\cosh_{4   \omega\chi} )+\cosh_{4 \chi  \omega} -1\right) \,,
	\\
	{(\Delta (\hat{V}_\chi\hat{H}))_{\text{C}}^2} + 	\omega^2 {\langle \hat{C}_\chi \rangle_{\text{C}}^2} &=\frac{v^2 \omega^2}{4} \left[ 4  | \alpha | ^4  \cosh ^2_{2   \omega\chi} \cos ^2_{2 \vartheta} +   \left(2 |\alpha|^2 (\cosh_{2   \omega\chi}\sin_{2 \vartheta}  +\sin_{2   \omega\chi} )+\sinh_{2   \omega\chi} \right)^2 \right] \,,
\end{align*}
%\end{widetext}
which can only be equal if $v \omega |\alpha|^2 \cosh_{2\omega \chi} =0$. Since we exclude the trivial cases with vanishing GFT parameters and with $\alpha=0$ (which would reduce a coherent state $|\alpha\rangle$ to the vacuum $|0\rangle$), there is no $\chi$ for which the Robertson--Schrödinger relation is saturated. Conversely, for squeezed states we find that the uncertainty principle is minimised at all times:
%\begin{widetext}
\begin{equation*}\label{RSSSchi}
	\begin{aligned}
		{(\Delta \hat{V}_\chi )_{\text{S}}^2}{(\Delta \hat{H})_{\text{S}} ^2} &= {(\Delta (\hat{V}_\chi\hat{H}))_{\text{S}}^2} + 	\omega^2 {\langle \hat{C}_\chi \rangle_{\text{S}}^2}\\ 
		&= \frac{v^2 \omega ^2}{16}  \Big[\sinh ^2_{2   \omega\chi}  \left(\sinh ^4_{2 r} \sin ^2_{2 \psi} +4 \cosh ^2_{2 r}\right)+\cosh ^2_{2   \omega\chi}  \left(4 \sinh ^2_{2 r} \sin ^2_\psi +\sinh ^2_{4 r} \cos ^2_\psi\right) \\ & \qquad\qquad  
		+ \frac{1}{4}   \sinh_{4   \omega\chi}  \sin_\psi \left(8 \sinh ^3_{2 r} \cosh_{2 r} \cos_{2 \psi}+6 \sinh_{4 r}+\sinh_{8 r}\right)\Big]\,.
	\end{aligned}
\end{equation*}
%\end{widetext}

\begin{figure}[ht]
	\begin{center}
		\includegraphics[width=.32\textwidth]{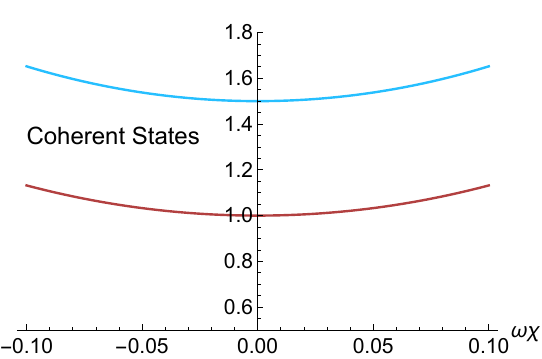}
		%\hspace{1cm}
		\includegraphics[width=.32\textwidth]{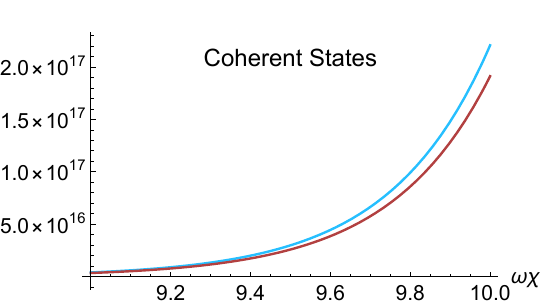}
		%	\hspace{1cm}
		\includegraphics[width=.32\textwidth]{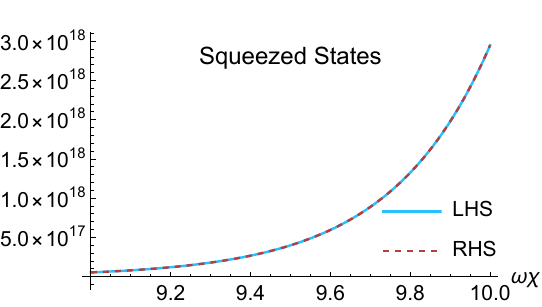}
	\end{center}
	\caption{\small Left-hand side (LHS) and right-hand side (RHS) of the RS inequality \eqref{RSprinciple} for coherent ($\alpha=1$) and squeezed ($z=1$) states, setting $v=1$. The first two panels show that coherent states do not saturate the inequality at early or late times; the last panel shows that squeezed states saturate \eqref{RSprinciple} at all times. 
	}
	\label{RUCSS}
\end{figure}

In other words, $\chi$-evolution does not change the statement of whether the RS uncertainty principle is saturated. Figure \ref{RUCSS} shows this feature for some state parameters.

Other than studying generic intermediate times, we can in particular also evaluate the large-volume limits \eqref{rs1} and \eqref{rs2} using the quantities in table \ref{TabCohSqu}. We find
%\begin{widetext}
\begin{align*}%\label{csrsl}
	\frac{(\Delta \hat{V}_\chi )_{\text{C}}^2}{\langle \hat{V}_\chi \rangle_{\text{C}}^2}\frac{(\Delta \hat{H})_{\text{C}} ^2}{\langle \hat{H} \rangle_{\text{C}}^2} \quad &\overset{\chi\rightarrow\pm\infty}{\longrightarrow}\quad \frac{ (2 |\alpha|^2+1)[1 +4|\alpha|^2(1\pm \sin_{2\vartheta})]}{|\alpha|^4 \cos_{2\vartheta}^2(1+2|\alpha|^2(1\pm\sin_{2\vartheta}))^2 } \,,
	\\%\label{csrsr}
	\frac{(\Delta (\hat{V}_\chi\hat{H}))_{\text{C}}^2}{\langle \hat{V}_\chi \rangle_{\text{C}}^2 \langle \hat{H} \rangle_{\text{C}}^2} + 	\omega^2 \frac{\langle \hat{C}_\chi \rangle_{\text{C}}^2}{\langle V_\chi \rangle_{\text{C}}^2 \langle \hat{H} \rangle_{\text{C}}^2} \quad &\overset{\chi\rightarrow\pm\infty}{\longrightarrow}\quad 
	\frac{1+4|\alpha|^2\left(2 |\alpha|^2+1\right)(1\pm\sin_{2\vartheta}) }{|\alpha|^4 \cos_{2\vartheta}^2(1+2|\alpha|^2(1\pm\sin_{2\vartheta}))^2} \,,
\end{align*}
%\end{widetext}
and
\begin{align*}%\label{ssrsl}
	\frac{(\Delta \hat{V}_\chi )_{\text{S}}^2}{\langle \hat{V}_\chi \rangle_{\text{S}}^2}\frac{(\Delta \hat{H})_{\text{S}} ^2}{\langle \hat{H} \rangle_{\text{S}}^2}  \; &\overset{\chi\rightarrow\pm\infty}{\longrightarrow}\; 4 + 4 \text{csch}^2_{2 r} \sec^2_\psi \,,
	\\%\label{ssrsr}
	\frac{(\Delta (\hat{V}_\chi\hat{H}))_{\text{S}}^2}{\langle \hat{V}_\chi \rangle_{\text{S}}^2 \langle \hat{H} \rangle_{\text{S}}^2} + 	\omega^2 \frac{\langle \hat{C}_\chi \rangle_{\text{S}}^2}{\langle V_\chi \rangle_{\text{S}}^2 \langle \hat{H} \rangle_{\text{S}}^2} \;&\overset{\chi\rightarrow\pm\infty}{\longrightarrow}\;  4 + 4 \text{csch}^2_{2 r} \sec^2_\psi  \,,
\end{align*}
showing confirmation of the above statements for both classes of states.

\begin{figure}[ht]
	\begin{center}
		\includegraphics[width=.32\textwidth]{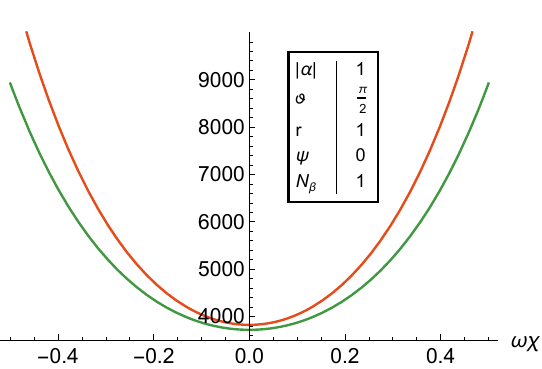}
		\hspace{1cm}
		\includegraphics[width=.35\textwidth]{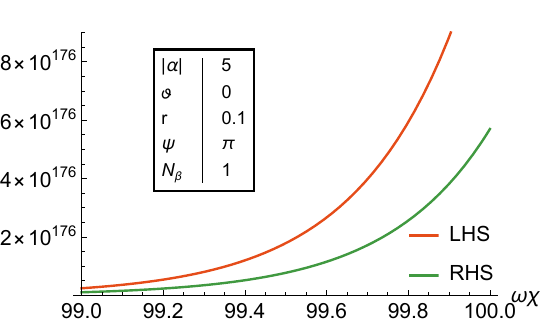}
	\end{center}
	\caption{\small Left-hand side (LHS) and right-hand side (RHS) of the RS principle \eqref{RSprinciple} for Gaussian states, setting $v=1$. The inequality is not saturated at early or late times regardless of the choice of parameters.}
	\label{RUGSS}
\end{figure}

In a similar fashion, one can also deal with the more general Gaussian states \eqref{GS}. As with coherent states, one finds that Gaussian states do not minimise the RS principle at any $\chi$. For $\chi=0$ one can quickly read off from the results in section \ref{GaussSection} that the right-hand side and the left-hand side of the inequality do not match, namely
%\begin{widetext}
\begin{align*}%\label{RSGS0}
	{(\Delta \hat{V} )_{\text{G}}^2}{(\Delta \hat{H})_{\text{G}} ^2} &=  \frac{v^2 \omega ^2}{32}  \left(4 |\alpha|^2 \mathcal{B} (\cosh_{2 r}+\mathcal{F}_+ \sinh _{2 r})+\mathcal{B}^2 \cosh_{4 r}-1\right) \times  \\ 
	& \quad\left[8 |\alpha|^2 \mathcal{B} (\cosh_{2 r}+\mathcal{F}_- \sinh_{2 r})	+\mathcal{B}^2 \left(2 \sinh ^2_{2 r} \cos _{2 \psi} +\cosh_ {4 r}+1\right)+2\right] \,,
	\\
	{(\Delta (\hat{V}\hat{H}))_{\text{G}}^2} + 	\omega^2 {\langle \hat{C} \rangle_{\text{G}}^2} & = \frac{v^2 \omega ^2}{16}  \Big[\mathcal{B}^2 \left(4 |\alpha|^2 ( \cosh_{2 r}\cos _{2 \vartheta }+\sinh_{2 r} \cos_\psi )+\mathcal{B} \sinh_{4 r} \cos_\psi \right)^2\\ & \qquad  \qquad+4 \left(2 |\alpha |^2 \sin_{2 \vartheta} +\mathcal{B} \sinh_{2 r} \sin_\psi \right)^2\Big] \,,
\end{align*}
%\end{widetext}
where $\mathcal{B}=  \coth_{\beta/2}$ and $\mathcal{F}_\pm =\cos_{2 \vartheta } \cos_\psi \pm \sin_{2 \vartheta } \sin_\psi$. As with any other states, one can also use the time-dependent expressions \eqref{chivar} and \eqref{chicovar} to compute the behaviour of the uncertainty principle under time evolution. Even though such results can be calculated analytically, we do not report the (lengthy) Gaussian state expressions because they are not insightful; we refer the reader to figure \ref{RUGSS} instead. On the other hand, we show explicitly that the inequality is not saturated for $\chi \rightarrow \pm \infty$. Using again the convenient large-volume limits \eqref{rs1} and \eqref{rs2}, one can compare the product of the (late-time) volume and Hamiltonian relative uncertainties
%\begin{widetext}
\begin{align*}
	&\frac{(\Delta \hat{V}_\chi )_{\text{G}}^2}{\langle \hat{V}_\chi \rangle_{\text{G}}^2}  \quad \overset{\chi\rightarrow\pm\infty}{\longrightarrow}\quad
	2- \frac{8 | \alpha | ^4 (\sin_{2 \vartheta} \pm1)^2}{  \left[ 2| \alpha |^2 (\sin_{2 \vartheta} \pm 1) \pm \mathcal{B} (\cosh_{2 r}\pm\sinh_{2 r} \sin_\psi )\right]^2}\,, \\
	&	\frac{(\Delta \hat{H})_{\text{G}} ^2}{\langle \hat{H} \rangle_{\text{G}}^2} \qquad=\qquad \frac{8 \mathcal{B} |\alpha |^2 (\cosh_{2 r}+\mathcal{F}_- \sinh_{2 r})+\mathcal{B}^2 \left(2 \sinh ^2_{2 r}\cos_{2 \psi}+\cosh_{4 r}+1\right)+2}{2 \left(2 |\alpha|^2 \cos_{2 \vartheta}+\mathcal{B} \sinh_{2 r} \cos_\psi \right)^2}\,,
\end{align*}
with
\begin{equation*}
	\begin{aligned}
		&	\frac{(\Delta (\hat{V}_\chi\hat{H}))_{\text{G}}^2}{\langle \hat{V}_\chi \rangle_{\text{G}}^2 \langle \hat{H} \rangle_{\text{G}}^2} +\omega^2 \frac{\langle \hat{C}_\chi \rangle_{\text{G}}^2}{\langle V_\chi \rangle_{\text{G}}^2 \langle \hat{H} \rangle_{\text{G}}^2} \quad   \overset{\chi\rightarrow\pm\infty}{\longrightarrow} \quad \frac{4}{\left(2 |\alpha|^2 \cos_{2 \vartheta }+\mathcal{B} \sinh_{2 r} \cos_\psi \right)^2} \: +  \\& \quad + \frac{\mathcal{B}^2 \left[4 |\alpha|^2 (\sinh_{2 r}
			(\cos_\psi \pm \sin_{2\vartheta + \psi})+ \cosh_{2 r}\cos_{2 \vartheta })+\mathcal{B} \left(\sinh_{4 r} \cos_\psi \pm \sinh ^2_{2 r} \sin _{2 \psi}  \right)\right]^2}{\left(2 |\alpha |^2 \cos_{2 \vartheta} +\mathcal{B} \sinh_{2 r}\cos_\psi \right)^2 \left[2 |\alpha  |^2 (\sin_{2 \vartheta} \pm1)+\mathcal{B} (\sinh_{2 r} \sin_\psi \pm\cosh_{2 r})\right]^2}\,,
	\end{aligned}
\end{equation*}
%\end{widetext}
to see that the saturation of the RS relation does not occur at late times. Since the complexity of Gaussian states makes these expressions somewhat intransparent, we report in figure \ref{RUGSS} a graphical demonstration of the exact time evolution for some state parameters. 

\section{Gaussian states and thermofield formalism}\label{GaussThermoApp}

In this appendix we present the features of Gaussian states that lead to the definition \eqref{GS}, as well as the tools used to obtain all the results of section \ref{GaussSection}. We start by means of a pivotal result, originally investigated in \cite{Schumaker,*MaRhodes,*MaRhodesMulti} (see also \cite{TextSerafini,Lucas} for modern perspectives), which states that when dealing with second-order bosonic Hamiltonians, the most general Gaussian state can always be expressed as three types of unitary operators acting on the thermal state $\hat{\rho}_\beta$ \eqref{thermalstate} (or on the Fock vacuum $|0\rangle$ for pure Gaussian states). These so-called ``fundamental Gaussian unitaries'' are the squeezing, displacement and rotation operators
\begin{equation}\label{FGU}
	\hat{S}(z) = e^{\frac{1}{2}(z\hat{a}^{\dagger 2}  - \bar{z}\hat{a}^2)}\,, \qquad \hat{D}(\alpha) = e^{\alpha \hat{a}^{\dagger}- \bar{\alpha}\hat{a}}\,,\qquad \hat{R}(\phi) = e^{{\rm i} \phi \hat{a}^\dagger \hat{a}} \,,
\end{equation}
where $(z,\alpha) \in \mathbb{C}$, $z = r e^{{\rm i}\psi}$ and $(r, \psi, \phi) \in \mathbb{R}$. 

To prove that a generic Gaussian state can be taken to be of the form \eqref{GS}, one can first notice that the operators in \eqref{FGU} satisfy
\begin{equation}\label{comp}
	\begin{aligned}
		\hat{R}(\phi_1)\hat{R}(\phi_2) &= \hat{R}(\phi_1+\phi_2)  \,,\\
		\hat{D}(\alpha_1)\hat{D}(\alpha_2) &= e^{\frac{1}{2}(\overline{\alpha_2}\alpha_1-\alpha_2 \overline{\alpha_1})}\hat{D}(\alpha_1+\alpha_2) \,,\\
		\hat{S}(z_1)\hat{S}(z_2) &=e^{{\rm i} \phi/2} \hat{S}(z_3) \hat{R}(\phi) \,,
	\end{aligned}
\end{equation}
where, defining $t_a = \tanh(r_a) \exp({\rm i}\psi_a)$, $\phi$ and $z_3$ are determined by $	e^{{\rm i} \phi} = \frac{1+t_1 \overline{t_2}}{|1+t_1 \overline{t_2}|}$ and $t_3 = \frac{t_1+t_2}{1+\overline{t_1}t_2}$. Moreover, one finds that operators of different types ``compose in a closed way'', namely
\begin{align}
	\hat{S}^\dagger(z)\hat{D}(\alpha)\hat{S}(z) &= \hat{D} (\alpha\cosh_r  - \bar{\alpha}e^{{\rm i}\psi}\sinh_r) \,, \label{closerules1} \\
	\hat{R}^\dagger(\phi)\hat{S}(z)\hat{R}(\phi) &= \hat{S} (e^{-2 {\rm i} \phi} z) \,, \label{closerules2}\\
	\hat{R}^\dagger(\phi)\hat{D}(\alpha)\hat{R} (\phi)&= \hat{D} (e^{-{\rm i} \phi} \alpha) \,. \label{closerules3}
\end{align}
Since the state parameters are arbitrary, properties \eqref{comp}--\eqref{closerules3} show that one can define a Gaussian state by acting on $\hat{\rho}_\beta$ with any number of the operators in \eqref{FGU}, in any order. Finally, we can see that it is no loss of generality to define Gaussian states without using the rotation operator. Using \eqref{closerules2} and \eqref{closerules3}, starting from an arbitrary definition of Gaussian states we could move the rotation operator until it acts on $\hat{\rho}_\beta$ (or on $|0\rangle$ in the case of zero temperature). These operations only change the free parameters $z$ and $\alpha$, which were arbitrary to being with, by a phase. But $\hat{R}(\phi)$ leaves both $\hat{\rho}_\beta$ and $|0\rangle$ invariant, and hence has no effect whatsoever.

Central to the computation of expectation values with (both pure and mixed) Gaussian states is the action of the displacement operator and squeezing operator on $\hat{a}$ and $\hat{a}^\dagger$:
\begin{alignat}{4}
	&	\hat{S}^\dagger (z)\hat{a} \hat{S}(z) &&= \cosh_r \hat{a} + e^{{\rm i}\psi} \sinh_r \hat{a}^\dagger\,, \qquad\qquad  &&\hat{S}^\dagger (z)\hat{a}^\dagger \hat{S}(z)  &&= \cosh_r \hat{a}^\dagger + e^{-{\rm i}\psi} \sinh_r \hat{a}\,, \label{sas}\\
	&	\hat{D}^\dagger (\alpha)\hat{a} \hat{D}(\alpha) &&= \hat{a}+\alpha \,,  &&\hat{D}^\dagger (\alpha)\hat{a}^\dagger \hat{D}(\alpha) &&= \hat{a}^\dagger+\bar{\alpha}\,. \label{dad}
\end{alignat}
From these one can see that expressions of the form $\hat{S}^\dagger(z)\hat{D}^\dagger  (\alpha) f(\hat{a},\hat{a}^\dagger) \hat{D} (\alpha)\hat{S}(z)$ are in general simpler than $\hat{D}^\dagger (\alpha)\hat{S}^\dagger (z) f(\hat{a},\hat{a}^\dagger )\hat{S}(z)\hat{D} (\alpha)$ for any function $f$ of the ladder operators. This is why, for instance, displaced squeezed states $|\alpha, z \rangle=\hat{D}(\alpha) \hat{S}(z) |0\rangle $ are usually adopted as pure Gaussian states instead of squeezed coherent states $|z, \alpha \rangle = \hat{S}(z) \hat{D}(\alpha) |0\rangle $ (the same applies to our general definition \eqref{GS}); the property \eqref{closerules1} effectively allows us to choose the most convenient ordering.

While by use of \eqref{sas} and \eqref{dad} one can obtain all the expressions of section \ref{GaussSection} using the density matrix $\hat{\rho}_\text{G}$ \eqref{GS} and the standard techniques for calculating traces, we outline here the \textit{thermofield formalism} as a very useful tool to extract the same results (in particular if one wishes to use a computational software, such as \textit{Mathematica}). This also allows us to link the present paper with the work of \cite{Isha_thermalGFT,*Isha_thermalGFT2,*IshaDaniele}, where such a formalism was adopted to introduce thermal effects in GFT.

Thermofield dynamics were introduced in \cite{thermofield0} (see \cite{ThermoBook} for a recent detailed treatment) as a formalism to link ensemble averages of statistical mechanics to expectation values computed with a temperature-dependent vacuum state, dubbed \textit{thermal vacuum} and denoted $|0_\beta\rangle$. In a nutshell, such a framework establishes a correspondence between a density matrix, which in our case is of thermal type $\hat{\rho}_\beta$ \eqref{thermalstate}, and the pure vector state $|0_\beta\rangle$, such that
\begin{equation}\label{thermofield}
	\tr \left( \hat{\rho}_\beta \hat{\mathcal{O}}\right) = \langle 0_\beta | \hat{\mathcal{O}}|0_\beta\rangle \,.
\end{equation} 
In \cite{thermofield0} it is shown that one can define a thermal vacuum satisfying \eqref{thermofield} only by enlarging the conventional Fock space. Specifically, one needs to double it by adding a fictitious system (denoted with a tilde) identical to the one under investigation. For simple theories such as our single-mode GFT cosmological models, this means introducing a new pair of bosonic ladder operators ($\hat{\tilde{a}}$, $\hat{\tilde{a}}^\dagger$) and constructing a second Fock space from a ``tilde vacuum'' $|\tilde{0}\rangle$ in the standard way, with
\begin{equation}
	\left[ \hat{{\tilde{a}}}, \hat{\tilde{a}}^\dagger \right] = 1 \,, \qquad \qquad  \hat{\tilde{a}}|\tilde{0}\rangle =0 \,.
\end{equation}
The tilde operators commute with the non-tilde operators as they live in distinct spaces. The next step in the thermofield formalism is to define a ``product vacuum'' in the doublet Hilbert space as a \textit{zero-temperature} ground state, from which one can construct product states in the usual manner,
\begin{equation}
	|0,\tilde{0}\rangle \equiv |0\rangle \otimes |\tilde{0}\rangle   \,, \qquad \qquad  \hat{a} |0,\tilde{0} \rangle = \hat{\tilde{a}} |0,\tilde{0} \rangle =0 \,,\qquad \qquad |n, \tilde{m}\rangle = \frac{(\hat{a}^\dagger)^n(\hat{\tilde{a}}^\dagger)^m}{\sqrt{n!}\sqrt{m!}}|0,\tilde{0}\rangle \,.
\end{equation}
Finally, the correspondence with thermal states is established by introducing temperature through a Bogoliubov transformation that mixes the real and fictitious systems:
\begin{equation}\label{thvacuum}
	|0_\beta\rangle = \hat{T}(\theta_\beta)  	|0,\tilde{0}\rangle \,,  \qquad \qquad \hat{T}(\theta_\beta) = e^{\theta_\beta(\hat{a}^\dagger\hat{\tilde{a}}^\dagger-\hat{a}\hat{\tilde{a}})} \,.
\end{equation}
The so-called thermalising operator $\hat{T}(\theta_\beta)$ is a two-mode squeezing operator and the real parameter $\theta_\beta$ encodes temperature in a way that will enable us to verify \eqref{thermofield}. One can easily check that the transformed, now \textit{temperature-dependent}, ladder operators
\begin{equation}\label{ladderbeta}
	\begin{aligned}
		\hat{a}_\beta &= \hat{T} (\theta_\beta)\hat{a} \hat{T}^\dagger (\theta_\beta)=\cosh_{\theta_\beta} \hat{a} - \sinh_{\theta_\beta} \hat{\tilde{a}}^\dagger \,,\\
		\tilde{\hat{a}}_\beta & = \hat{T} (\theta_\beta) \tilde{\hat{a}} \hat{T}^\dagger(\theta_\beta)= \cosh_{\theta_\beta} \hat{\tilde{ a}} - \sinh_{\theta_\beta} {\hat{a}}^\dagger \,,
	\end{aligned}
\end{equation}
indeed annihilate the state \eqref{thvacuum}
\begin{equation}
	\hat{a}_\beta|0_\beta\rangle =   \hat{\tilde{a}}_\beta| 0_\beta\rangle =0\,.
\end{equation}
The operators \eqref{ladderbeta} and their adjoints still satisfy the bosonic algebra because Bogoliubov transformations are canonical.  Therefore, one can construct a Fock space with respect to the $\beta$-dependent operators, where general states read
\begin{equation}
	|n, \tilde{m} ; \beta\rangle = \hat{T}(\theta_\beta) |n,\tilde{m}\rangle = \frac{(\hat{a}_\beta^\dagger)^n(\hat{\tilde{a}}_\beta^\dagger)^m}{\sqrt{n!}\sqrt{m!}}|0_\beta\rangle \,.
\end{equation}	

The association of the thermal vacuum $|0_\beta\rangle$ with a density matrix $\hat{\rho}_\beta$ as given in \eqref{thermalstate} is obtained by determining $\theta_\beta$ as a function of $\beta$. This is done by imposing the condition \eqref{thermofield} for the number operator so that, thanks to \eqref{Nbeta}, one has
\begin{equation}\label{thermomap}
	\frac{1}{e^{\beta } -1} = 	\tr \left(\hat{\rho}_\beta \hat{a}^\dagger\hat{a}\right) =  \langle 0_\beta |\hat{a}^\dagger\hat{a} |0_\beta\rangle = \sinh^2_{\theta_\beta}\,,
\end{equation}
where the right-hand side is computed using the thermal vacuum \eqref{thvacuum}. The fictitious system is understood as unphysical and is only introduced as a useful tool. We are not interested in expectation values of tilde operators; this helps to define a simple thermofield counterpart of Gaussian states.

Starting from the thermal vacuum $|0_\beta\rangle$, we can construct states by analogy with \eqref{CS} and \eqref{SS} and hence use displacement and squeezing operators to define general Gaussian states in the thermofield formalism. Indeed, we can generalise the GFT coherent thermal states of \cite{Isha_thermalGFT,*Isha_thermalGFT2,*IshaDaniele}, defining the thermofield analogue of $\hat{\rho}_\text{G} $ (cf.\ \eqref{GS}) as
\begin{equation}\label{thermalGauss}
%	\begin{aligned}
	|\alpha, z , \beta\rangle =	\hat{D}(\alpha) \hat{S}(z) \hat{T}(\theta_\beta) |0, \tilde{0} \rangle = \hat{D}(\alpha) \hat{S}(z)|0_\beta \rangle \,,
%		\end{aligned}
\end{equation}
which clearly reduces to a pure Gaussian state when $\beta\rightarrow\infty$ (or equivalently $\theta_\beta \rightarrow 0$). Since the thermalising operator \eqref{thvacuum} is of squeezing type, the order in which the operators appear in \eqref{thermalGauss} is in principle arbitrary (see above discussion) and was chosen for convenience. There is however a subtlety: while $\hat{D}(\alpha)$ and $\hat{S}(z)$ only act on the non-tilde sector, $\hat{T}(\theta_\beta)$ is a two-mode operator which mixes the physical and fictitious parts. Thus, choosing $\hat{T}(\theta_\beta)$ to be to the left of $\hat{D}(\alpha)$ and/or $\hat{S}(z)$ requires the introduction of $\hat{\tilde{D}}({\tilde{\alpha}})$ and/or $\hat{\tilde{S}}({\tilde{z}})$, with $\tilde{\alpha}= \bar{\alpha}$ and $\tilde{z} = \bar{z}$ (see \cite{ThermoBook} for details).\footnote{One can for example use the state $\hat{T}(\theta_\beta)	\hat{D}(\alpha) \hat{\tilde{D}}({\bar{\alpha}}) \hat{S}(z) \hat{\tilde{S}}({\bar{z}})   |0, \tilde{0} \rangle $ or any other alternative definition achieved by changing the operator ordering; they will all be related to $|\Psi_\text{G} ; \beta\rangle$ in \eqref{thermalGauss} via \eqref{closerules1}.} 

On the other hand, if the mixing between the physical and the fictitious sectors happens first (i.e., $\hat{T}(\theta_\beta)$ is to the right of every other operator as in \eqref{thermalGauss}), one is free to displace and squeeze only the non-tilde component of the state. Of course, one can also include the tilde operators $\hat{\tilde{D}}({\bar{\alpha}})$ and $\hat{\tilde{S}}({\bar{z}})$ acting from the left in \eqref{thermalGauss} (as done in \cite{Isha_thermalGFT,*Isha_thermalGFT2,*IshaDaniele} for coherent states), but this is not necessary; it would only be relevant if one were interested in the expectation values of tilde operators. In short, the state \eqref{thermalGauss} is the simplest thermofield analogue of the Gaussian density matrix \eqref{GS}.

The explicit computational convenience of the thermofield formalism comes from the following transformation rules of our GFT ladder operators
\begin{equation}\label{TaT}
	\begin{aligned}
		\hat{T}^\dagger(\theta_\beta) \hat{a} \hat{T}(\theta_\beta) &=  	\cosh_{\theta_\beta} \hat{a} + \sinh_{\theta_\beta} \hat{\tilde{a}}^\dagger						\,,	\\
		\hat{T}^\dagger(\theta_\beta) \hat{a}^\dagger \hat{T}(\theta_\beta) &= 		 \cosh_{\theta_\beta}  \hat{a}^\dagger + \sinh_{\theta_\beta} \hat{\tilde{a}} \,,
	\end{aligned}
\end{equation}
which were tacitly used to compute the right-hand side of \eqref{thermomap}. The transformations \eqref{TaT}, together with \eqref{sas} and \eqref{dad}, allow us to forget entirely about Gaussian density matrices and taking traces. This makes calculations very mechanical: one first simply computes expectation values with \eqref{thermalGauss} exactly as if using pure states; then, one makes use of the identification \eqref{thermomap} to map the thermofield results to the standard ($\beta$-dependent) results of the density matrix approach.

\section{Condensate states and physicality conditions}\label{AppCondensates}

In this appendix we focus on states that have been proposed in the algebraic approach to GFT \cite{GFTquantumST_Oriti,*ORITI2017235,Gielen_2016,GFTcosmoLONGpaper}, and study their properties in relation to the questions addressed in this paper. As mentioned in section \ref{algebraicsection}, the single-particle condensate state ubiquitously adopted in the literature,
\begin{equation}\label{cond1}
	|{\sigma }\rangle = \mathcal{N}_{{\sigma}} \exp\left(\int {\rm d} \chi \, {\sigma} (\chi) \hat{\varphi}^\dagger(\chi) \right)|0\rangle \,, \qquad \mathcal{N}_{{\sigma}} = \exp\left( -\frac{1}{2} \int {\rm d} \chi |{\sigma}(\chi)|^{2} \right) \,,
\end{equation}
with $\sigma(\chi)$ solving the classical equations of motion (cf.\ \eqref{volumeALG}), is an exact solution of the constraint \eqref{constraint} for a free GFT, as it is equivalent to \eqref{sigma}. After some regularisation (e.g., a cut-off in $\chi$) is adopted to make the state well-defined, this yields compelling cosmological effective dynamics \eqref{FriedmannSigma} as well as volume uncertainties \eqref{deltasigma} which are well-behaved if one considers smeared observables.

Next to \eqref{cond1}, two-particle (or dipole) condensate states
\begin{equation}\label{cond2}
	| {\xi} \rangle = \mathcal{N}_{{\xi}} \exp \left(\frac{1}{2} \int {\rm d} \chi \,  {\xi} (\chi) \hat{\varphi}^\dagger (\chi) \hat{\varphi}^\dagger (\chi)\right) |0\rangle\,, \qquad   \mathcal{N}_{{\xi}} = \exp\left( -\sum_{k=1}^\infty \frac{1}{4k} \delta(0) \int {\rm d} \chi |{\xi}(\chi)|^{2k} \right) 
\end{equation}
were also initially proposed for GFT cosmology \cite{GFTcosmoLONGpaper}, even though they were never used to obtain relational dynamics. Here we return to these states because of the connection with squeezed states (see section \ref{SSSection}, and in particular \eqref{map}), which makes them part of the Gaussian states family. Indeed, assuming one can make such two-particle states well-defined, dipoles \eqref{cond2} and squeezed-like states $\hat{\mathcal{S}}(\zeta)|0\rangle$ (with $\hat{\mathcal{S}}(\zeta)$ in \eqref{squeezingALG}) correspond to the same type of states and share the same properties. 

An important difference between the normalisation factor $ \mathcal{N}_{{\xi}}$ in \eqref{cond2} and the one given in \cite{GFTcosmoLONGpaper} is the appearance of a Dirac delta function $\delta(0)$ due to the commutator \eqref{algcom}, which does not appear for models without a matter scalar field. A cut-off in $\chi$, while required to make sense of the integrals in $\mathcal{N}_\xi$, is not enough to make the state well-defined; some other regularisation would be needed to deal with the $\delta(0)$. As in section \ref{SSSection}, we will proceed to study properties of dipole states assuming that a way of regularising \eqref{cond2} is found (and hence leaving formal $\delta(0)$ factors in our expressions). We shall see that, even assuming they exist, dipole states are not viable candidates as semiclassical states for the class of GFT models discussed in this paper, for a number of reasons.

Formally keeping divergences in all our expressions below, one can ``blindly'' follow the usual strategy to derive a cosmological scenario by focussing on the volume operator. Computing its expectation value with respect to \eqref{cond2}, one finds
\begin{equation}\label{volxi}
%	\begin{aligned}
	\langle\hat{V}(\chi)\rangle_\xi = v\frac{\langle\xi | \hat{\varphi}^\dagger(\chi) \hat{\varphi}(\chi) |\xi \rangle}{\langle \xi |\xi\rangle}= v \delta(0) \frac{|\xi(\chi)|^2}{1-|\xi(\chi)|^2}\,,
%	\end{aligned}
\end{equation}
where again the commutator \eqref{algcom} gives rise to a $\delta(0)$. Since the expectation value \eqref{volxi} is positive by construction, one concludes that the dipole function must satisfy the condition 
\begin{equation}\label{xil1}
	|\xi(\chi)|<1\,,
\end{equation}
similarly to what we observed in section \ref{SSSection}. Next, we shall find that for any $\xi(\chi)$, quantum fluctuations of these states are never small. To see this, one first computes the volume variance as
\begin{equation}
	\begin{aligned}
		(\Delta \hat{V}(\chi) )_\xi^2 &= v^2 \Big[\delta (0) \langle \hat{\varphi}^\dagger(\chi) \hat{\varphi}(\chi)\rangle_\xi  +  \langle \hat{\varphi}^{\dagger 2}(\chi) \hat{\varphi}^2(\chi)\rangle_\xi  -\langle \hat{\varphi}^\dagger(\chi) \hat{\varphi}(\chi)\rangle_\xi ^2 \Big] \\ &= v^2 \delta^2(0) \frac{2|\xi(\chi)|^2}{(|\xi(\chi)|^2-1)^2} \,,
	\end{aligned}
\end{equation}
so that the relative uncertainty reads
\begin{equation}\label{2overxi}
	\frac{(\Delta \hat{V}(\chi) )_\xi^2}{\langle\hat{V}(\chi) \rangle_\xi^2} = \frac{2}{|\xi(\chi)|^2}\,.
\end{equation}
Even though the $\delta(0)$ distributions fortuitously cancel (as already seen below \eqref{deltagaussbis}), such fluctuations can never be made small because of the condition \eqref{xil1}. This means that dipole states never become semiclassical according to our main criteria.

One can now turn to the task of finding the form of $\xi(\chi)$ using dynamical equations, but as anticipated for squeezed-like states in section \ref{algebraicsection}, this does not seem to be possible. First of all, the dipole state \eqref{cond2} is not exactly physical since it does not solve the constraint \eqref{constraint}. To be precise, using the property $ \hat{\varphi} (\chi) |  \xi  \rangle =  \xi   (\chi ) \hat{\varphi}^\dagger(\chi)|\xi \rangle $, from \eqref{constraint} one can obtain the condition
\begin{equation}\label{xicondition}
	\left(\partial^2_\chi - \omega^2 \right)\left( \xi (\chi) \hat{\varphi}^\dagger(\chi)\right) =0 \,,
\end{equation}
which cannot yield a solution for the dipole function $\xi(\chi)$. \eqref{xicondition} is  the analogue of \eqref{squnotphy} translated into the notation of dipole states via \eqref{map}.

One could then try to find $\xi(\chi)$ as an approximate solution from expectation values of the type \eqref{SDeq}. As already assessed in section \ref{algebraicsection} for more general states (cf.\ \eqref{averageGaussian}), the first Schwinger--Dyson equation does not help in this respect since dipoles have vanishing field expectation value. Climbing the tower of Schwinger--Dyson equations would entail setting, e.g., $\hat{\mathcal{O}} = \hat{\varphi}$ in \eqref{SDeq}, which gives
\begin{equation}
	\left(\partial^2_\chi - \omega^2 \right)   \langle \xi | \hat{{\varphi}} (\chi') 	  \hat{{\varphi}} (\chi)  | \xi  \rangle =0 \,.
\end{equation}
Calculating explicitly the correlation function $\langle \xi | \hat{{\varphi}} (\chi') 	  \hat{{\varphi}} (\chi)  | \xi  \rangle$ yields
\begin{equation}\label{mess1}
	\left(\partial^2_\chi - \omega^2 \right)   \left( \delta(\chi-\chi')  \frac{\xi(\chi)+\xi(\chi')|\xi(\chi)|^2}{1-|\xi(\chi')|^2 |\xi(\chi)|^2} \right)=\left(\partial^2_\chi - \omega^2 \right)   \left( \delta(\chi-\chi')\frac{\xi(\chi)}{1-|\xi(\chi)|^2} \right) =0 \,,
\end{equation}
where in the last step we used the relation $\delta(\chi-\chi')f(\chi,\chi')=\delta(\chi-\chi')f(\chi)$. To show the connection with squeezed-like states one last time, one can write the same Schwinger--Dyson equation using the state $|\zeta\rangle =\hat{\mathcal{S}}(\zeta)|0\rangle$ defined via \eqref{squeezingALG}, finding
\begin{equation}\label{mess2}
	\left(\partial^2_\chi - \omega^2 \right)   \left( \delta(\chi-\chi')\frac{\zeta(\chi)}{2|\zeta(\chi)|}\sinh(2|\zeta(\chi)|) \right)=0\,.
\end{equation}
Unfortunately, it seems that no nontrivial solution to either \eqref{mess1} or \eqref{mess2} exists. Schwinger--Dyson equations of higher order would be even harder to solve. In short, two-particle states such as squeezed-like states or dipoles \eqref{cond2} do not seem to be suitable candidate states for the GFT models under investigation. Specifically, while they are clearly not semiclassical (cf.\ \eqref{2overxi}), it is also not clear whether one can tackle the problem of finding a condition for the function $\xi(\chi)$ so to extract dynamics (in the same way that $\sigma(\chi)$ gave rise to \eqref{FriedmannSigma}), in any meaningful way.

These observations are in conflict with some results of \cite{GFTcosmoLONGpaper}, where a detailed account on dipole states can be found. In particular, in the specific case of a GFT coupled to a matter scalar field $\chi$, the condition on the dipole function is \textit{not} given by the classical GFT equation of motion \eqref{eom},
\begin{equation}\label{wrong}
	\left(\partial^2_\chi - \omega^2 \right)  \xi(\chi) =0\,,
\end{equation}
as was claimed in \cite{GFTcosmoLONGpaper}. Interestingly, plugging the solution of \eqref{wrong} (i.e., $\xi(\chi) = \alpha e^{\omega \chi} + \beta e^{-\omega \chi}$) into the volume expectation value \eqref{volxi} one finds new effective cosmological dynamics, where the volume diverges as $\langle V(\chi)\rangle_\xi \sim  (\chi-\chi_0)^{-1}$ in a finite relational time $\chi=\chi_0$.\footnote{If $(\alpha,\,\beta)\in \mathbb{R}$, one finds $\langle V(\chi)\rangle_\xi \sim \frac{v (1-4 \alpha\beta)^{-1/2}}{ 2 \omega(\chi-\chi_0)}$, with time of divergence given by $\omega \chi_0 = \ln\left( \frac{ 1+\sqrt{1-4 \alpha\beta}}{2 \alpha} \right)$.} In particular, assuming $\alpha$ and $\beta$ are small (so to satisfy \eqref{xil1}) and real for simplicity, one has the effective Friedmann equation
%\begin{widetext}
\begin{equation}\label{DE}
	\left(\frac{1}{\langle \hat{V}(\chi)\rangle_\xi } \frac{{\rm d} \langle \hat{V}(\chi)\rangle_\xi}{{\rm d} \chi}\right)^2 = 4 \omega ^2 \left( C_1  -\frac{v C_2 }{\langle \hat{V}(\chi)\rangle_\xi}+\frac{C_3}{v}\langle \hat{V}(\chi)\rangle_\xi +\frac{C_4}{v^2}\langle \hat{V}(\chi)\rangle_\xi^2 \right)\,,
\end{equation} 
%\end{widetext}
where $C_1=  1-12 \alpha\beta$, $C_2= 4 \alpha\beta $, $C_3 = 2-12 \alpha\beta$ and $C_4 = 1-4 \alpha\beta$. Note that the last two terms in \eqref{DE} \textit{effectively} behave like matter components with equations of state typical of ``dust'' and ``dark energy'', respectively. To see this explicitly, we recall that in classical cosmology the relational Friedmann equation for the volume as a function of $\chi$ takes the form \cite{Gielen_2020,OritiPang}
\begin{equation}\label{classicalDE}
	\left(\frac{1}{ {V}(\chi) } \frac{{\rm d}  {V}(\chi)}{{\rm d} \chi}\right)^2 = \sum_i A_i V(\chi) ^{-w_i+1}\,,
\end{equation} 
where $A_i$ are constants and $i$ labels the different types of perfect fluids (with equations of state $p_i=w_i \rho_i$) in the Universe. Comparing \eqref{DE} and \eqref{classicalDE} one can readily find the ``effective equation of state parameters'' $w_3 =0$ and $w_4 =-1$. Note that a term $\sim1/\langle \hat{V}(\chi)\rangle_\xi^2$, usually characterising effective Friedmann equations coming from GFT, is absent in the ``dipole cosmological dynamics'' \eqref{DE}. While this path might have some interesting implications from a phenomenological point of view, it is not clear why one should assume \eqref{wrong} to begin with.

To shed some light on the more general question of finding physical states, we conclude by deriving the general solution to the constraint \eqref{constraint},  or
\begin{equation}\label{Kphi}
	\left(\partial^2_\chi - \omega^2 \right)	\hat{\varphi} (\chi) |  \Phi  \rangle =  0 \,.
\end{equation}
Any element $|\Phi\rangle $ of the Fock space can be written as
\begin{equation}\label{genericstate}
	|\Phi\rangle = \sum_{n=0}^\infty \int {\rm d} \chi_1 \dots {\rm d} \chi_n \, f_n (\chi_1,\dots \chi_n) \hat{\varphi}^\dagger(\chi_1) \dots\hat{\varphi}^\dagger(\chi_n) |0\rangle \,,
\end{equation} 
where the functions $f_n$ are totally symmetric under exchange of their arguments. Substituting this form into \eqref{Kphi} and using the commutator \eqref{algcom}, it follows that we would need
\begin{equation}
	\sum_{n=0}^\infty n \int  {\rm d} \chi_1 \dots {\rm d} \chi_{n-1} \, \left(\partial^2_\chi - \omega^2 \right)  f_n(\chi,\chi_1,\dots, \chi_{n-1})  \hat{\varphi}^\dagger(\chi_1) \dots\hat{\varphi}^\dagger(\chi_{n-1}) |0\rangle = 0 \,,
\end{equation}
which is true if and only if
\begin{equation}\label{conditions}
	\int  {\rm d} \chi_1 \dots {\rm d} \chi_{n-1}  \left(\partial^2_\chi - \omega^2 \right)  f_n(\chi,\chi_1,\dots, \chi_{n-1})  \hat{\varphi}^\dagger(\chi_1) \dots\hat{\varphi}^\dagger(\chi_{n-1}) =0 \qquad \forall n\,.
\end{equation}
If such equations are satisfied by some $f_n$, then the state $|\Phi\rangle$ is physical. For example, forgetting about normalisation, the coherent state \eqref{cond1} corresponds to \eqref{genericstate} with
\begin{equation}\label{fnsingle}
	f_n(\chi_1,\dots,\chi_n) = \frac{1}{n!} \sigma(\chi_1)\dots \sigma(\chi_n) \,.
\end{equation}
The conditions \eqref{conditions} for the first few $n$'s read
%\begin{widetext}
\begin{equation}\label{csf}
	\begin{aligned}
		(\partial_\chi^2 - \omega^2) \sigma (\chi) &=0\,, \qquad \qquad n=1\,,\\
		\int {\rm d} \chi_1\, \sigma(\chi_1)\, \hat{\varphi}^\dagger(\chi_1) 	(\partial_\chi^2 - \omega^2) \sigma(\chi)  &=0\,, \qquad \qquad n=2\,,\\
		\int {\rm d} \chi_1\, {\rm d} \chi_2\,  \sigma( \chi_1) \sigma (\chi_2)\,\hat{\varphi}^\dagger(\chi_1) \hat{\varphi}^\dagger(\chi_2)   	(\partial_\chi^2 - \omega^2) \sigma (\chi)&=0 \,, \qquad \qquad n=3\,,\\
		& \vdots
	\end{aligned}
\end{equation}
%\end{widetext}
Clearly all these conditions are met if $\sigma(\chi)$ satisfies the classical GFT equation of motion \eqref{eom}. {Note} that the constants $1/n!$ in \eqref{fnsingle} do not play a role in obtaining \eqref{csf}, so \eqref{fnsingle} can straightforwardly be generalised to $f_n(\chi_1,\dots,\chi_n) = c_n \sigma(\chi_1)\dots \sigma(\chi_n)$, where $c_n$ are generic coefficients. In other words, one can define a slightly more general class of exact physical states,
\begin{equation}\label{Fsigma}
	|F_\sigma\rangle = F\left( \int {\rm d} \chi \, \sigma(\chi) \hat{\varphi}^\dagger(\chi) \right) |0\rangle \,,
\end{equation} 
where $F$ can be any function that can be expressed in a power series, generalising the previously used exponential. However, \textit{all} these (physical) states are non-normalisable regardless of $F$; even the simple one-particle state $	|\, \vcenter{\hbox{\adjincludegraphics[width=.025\textwidth]{tetstate.pdf} }}\rangle= \int {\rm d} \chi \, \sigma(\chi) \hat{\varphi}^\dagger(\chi)|0\rangle$, for example, would require a cut-off as $\langle \, \vcenter{\hbox{\adjincludegraphics[width=.025\textwidth]{tetstate.pdf} }} |\, \vcenter{\hbox{\adjincludegraphics[width=.025\textwidth]{tetstate.pdf} }} \rangle = \int {\rm d} \chi |\sigma(\chi)|^2 $. It seems the only regular physical state of the theory (when defined without cut-off) is the Fock vacuum.

As a last example, we can use the general expressions \eqref{conditions} to confirm that the two-particle condensate state \eqref{cond2} cannot be a physical state. Again forgetting about the normalisation, \eqref{genericstate} reduces to a dipole state by choosing $f_n(\chi_1, \dots,\chi_n) = 0$ for $n=2m+1 $ and
\begin{equation}	
	f_n(\chi_1, \dots,\chi_n) = \frac{1}{2^m m!} \delta(\chi_1-\chi_{m+1}) \dots \delta(\chi_m-\chi_{2m}) \xi(\chi_{m+1}) \dots \xi (\chi_{2m})\quad \text{for} \quad n=2m\,,
\end{equation}
where $m\in \mathbb{N}$. The state would be physical if the conditions \eqref{conditions} are satisfied. However, excluding the trivial (odd) ones, the first few read
%\begin{widetext}
\begin{equation}\label{csf2}
	\begin{aligned}
		(\partial_\chi^2 - \omega^2)\left( \xi (\chi) \hat{\varphi}^\dagger(\chi)\right) &=0\,,\qquad \qquad n=2\,,\\
		\int {\rm d} \chi_1 \,\xi (\chi_1) (\hat{\varphi}^\dagger(\chi_1)) ^2	(\partial_\chi^2 - \omega^2) \left( \xi (\chi) \hat{\varphi}^\dagger(\chi)\right)    &=0 \,,\qquad \qquad n=4\,,\\
		\int {\rm d} \chi_1 {\rm d} \chi_2 \,  \xi ( \chi_1) \xi  (\chi_2)(\hat{\varphi}^\dagger(\chi_1) )^2(\hat{\varphi}^\dagger(\chi_2) )^2  	(\partial_\chi^2 - \omega^2) \left( \xi (\chi) \hat{\varphi}^\dagger(\chi)\right)  &=0\,, \qquad \qquad n=6\,,\\
		&\vdots
	\end{aligned}
\end{equation}
%\end{widetext}
which reduce to the condition \eqref{xicondition}. 

In conclusion, it seems that only states built from iterations of the same single creation operator $\int {\rm d} \chi \, \sigma(\chi) \hat{\varphi}^\dagger(\chi)$, generically given by \eqref{Fsigma} and in particular including the coherent state \eqref{cond1} or equivalently \eqref{sigma}, are exact solutions of \eqref{Kphi}, and hence physical states. To the best of our knowledge, no other (nontrivial) state can be found such that the infinitely many conditions \eqref{conditions} are satisfied. This question needs to be investigated more in future work.

%\end{widetext}

\twocolumngrid
 
%%%%%%%%%%%%%%%%%%%%%%%  Bibliography %%%%%%%%%%%%%%%%%%%%%%%
\let\oldaddcontentsline\addcontentsline
\renewcommand{\addcontentsline}[3]{}
\bibliography{gaussGFT}
\let\addcontentsline\oldaddcontentsline
%%%%%%%%%%%%%%%%%%%%%%%%%%%%%%%%%%%%%%%%%%%%%%%%%%%%%%%%%%%%%%%

\end{document}